-------------------------------------------------------------------------
\documentstyle[12pt,a4wide]{article}
\relax

\newcommand{\beq} {\begin{equation}}
\newcommand{\eeq} {\end{equation}}
\newcommand{\bea}{\begin{eqnarray}}
\newcommand{\eea}{\end{eqnarray}}
\newcommand{\kk} {{\cal K}^2}

\newcommand{\lag} {{\cal L}}

\newcommand{\eps} {\varepsilon}
\newcommand{\opi} {{\bf I}}
\newcommand{\opj} {{\bf J}}

\newcommand{\xiq} {\xi^2}
\newcommand{\Pc} {\Phi_{cl}}
\newcommand{\Dt} {\tilde{\cal D}}
\newcommand{\Ft} {\tilde{F}}
\newcommand{\nn} {\nonumber\\}
\newcommand{\ds}{\displaystyle}

\begin{document}
\begin{titlepage}
\begin{flushright}
DO-TH-93/19 \\
July 1993
\end{flushright}

\vspace{20mm}
\begin{center}
{\Large
Quantum Fluctuations around the Electroweak Sphaleron}
\vspace{10mm}

{\large  J. Baacke\footnote{e-mail
baacke@het.physik.uni-dortmund.de} and S. Junker} \\
\vspace{15mm}

{\large Institut f\"ur Physik, Universit\"at Dortmund} \\
{\large D -- 44221 Dortmund, Germany}
\vspace{40mm}

\bf{Abstract}

We present an analysis of the quantum fluctuations around the
electroweak sphaleron and calculate the associated determinant
which gives the 1--loop correction to the sphaleron transition rate.
The calculation differs in various technical aspects from a previous
analysis by Carson et al. so that it can be considered as
independent. The numerical results differ also -- by several orders
of magnitude -- from those of this previous analysis; we find that
the sphaleron transition rate is much less suppressed than found
previously.
\end{center}
\end{titlepage}
\section{Introduction}
The electroweak theory is known
\cite{Ma,KlMa} for quite some time to possess a topologically
nontrivial solution which describes a saddlepoint between two
topologically distinct vacua. The recent interest in this solution
has been centered around its possible r\^ole in generating baryon
number violating processes in the early universe or even at
accelerator energies \cite{KuRuSha}--\cite{Ri}.
The rate of sphaleron transitions in the range of temperatures
$M_W(T) \ll   T \ll   M_W(T)/\alpha_w    $ has been
derived on the basis of the work of Langer \cite{La}
and Affleck \cite{Af} by Arnold and McLerran \cite{ArMcL}. It
is given by
\beq
\Gamma=\frac{\omega_{-}}{2 \pi}{\cal N} e^{-E_{cl}/T}\kappa.
\eeq
Here $\omega_{-}$ is the absolute value of the eigenvalue of the
unstable mode, the prefactor ${\cal N} $ refers to normalizations
introduced by the translation and rotation zero modes and is
given in detail below. $E_{cl}$
is the classical sphaleron energy and the factor $\kappa$ takes
account of the quantum fluctuations of the sphaleron.
It is given by
\beq
\kappa = {\rm Im} (\frac{\det \Delta_{FP}^S
\det \Delta_{gf}^0}{\det \Delta_{FP}^0
\det \Delta^{'S}_{gf}})^{1/2}
\eeq
where the symbols $\Delta$ denote the small fluctuation
operators. They are obtained by expanding the gauge fixed action
(gf) and the  Fadeev--Popov action (FP) evaluated around the
sphaleron (S) and the vacuum ($0$), respectively. If the fluctuation
operators are diagonalized the determinants are formally given by
the product of the squared eigenfrequencies $\lambda_{\alpha}^2$.
 The determinant $\det \Delta^{'S}_{gf}$  of the gauge fixed action
is to be evaluated without the zero modes and with the
eigenvalue of the unstable mode, $\lambda_{-}^2=-\omega_{-}^2$
replaced by its absolute value.

The first evaluation of $\kappa$  by Akiba, Kikuchi and Yanagida
\cite{AkKiYa} was restricted to the three lowest partial waves.
The authors concluded from their results that the inclusion
of this correction did not suppress the sphaleron transition
rate. Of course they considered their conclusion as
tentative and to be checked by a more precise evaluation.
Subsequently Carson and McLerran \cite{cmcl1} evaluated $\kappa$
in an approximation scheme (referred to as DPY in the following)
developed by Diakonov, Petrov and Yung
\cite{DiPeYu}, a complete exact numerical evaluation
was presented first by Carson et al. \cite{cmcl2}
(referred to as CLMW in the following).
The results of this exact calculation differ significantly
from the DPY approximation and from a perturbative estimate.
It is therefore of interest to repeat this exact evaluation
and this is -- besides a general analysis of the
fluctuation Lagrangean and its partial wave reduction --
the subject of our investigation.

Our investigation departs from the one of CLMW
in various points.
While we use the background gauge as CLMW,
we use another angular momentum basis and a different
scheme for evaluating the determinant.

The analysis of the small fluctuations around the sphaleron
requires a partial wave decomposition with respect to the
quantum number $\vec{K} = \vec{J} + \vec{I}$ ($K$--spin).
Our present work is based on the analysis of Ref. \cite{BaLa}
where the sphaleron stability was investigated.
The small fluctuation equations obtained there had to be modified
however. While that investigation was performed in
the $A_0=0$ gauge we use here the background gauge. So
a gauge fixing term had to be added and also we had to
construct the small fluctuation Lagrangean for the Fadeev--Popov
modes. The evaluation of the determinant has been performed using
the Euclidean Green function technique in analogy to some recent
investigations of one of the authors (J.B.) \cite{baanum}. The
results of our calculation
have already been communicated previously in a short
version \cite{BaJu}. Here we present an extensive
version including the complete partial wave analysis.

The paper is organized as follows. We present the basic relations
for the electroweak theory and the sphaleron solution in section 2.
The small fluctuation expansion and its partial wave analysis
are discussed in sections 3 and 4, respectively.
The explicit equations of motion which constitute our first main
result are presented in Appendix A.
In section 5 we collect the main formulae for the 1--loop correction
to the sphaleron transition rate and define more precisely the
zero mode prefactors and the fluctuation determinant, some material
being deferred to Appendix B.
In section 6 we express the fluctuation
determinant in terms of Euclidean Green functions, a formulation
that is used as the basis of our numerical evaluation. Sections 7
and 8 contain the discussion of renormalization and of the treatment
of the zero and unstable modes.
Some aspects of the numerical calculations and the
results are presented in section 9 and in Appendix C.
\setcounter{equation}{0}
\section{Basic Relations }
\label{grundtheo}
The action of a pure SU(2) gauge theory with minimal Higgs sector is
given in Minkowski space as
\beq
  S = \int d^4\bar{x}\,\bar{\lag}(\bar{x}),
\eeq
where the Lagrangean density is given by
\beq
  \bar{\lag} = - \frac{1}{4} \bar{F}_{\mu\nu}^a \bar{F}^{\mu\nu a}
         + \left(\bar{{\cal D}}_\mu\bar{\Phi}\right)^\dagger
           \left(\bar{\cal D}^\mu\bar{\Phi}\right)
         - \lambda \left(\bar{\Phi}^\dagger\bar{\Phi}
         - \frac{1}{2}v^2\right)^2.
\end{equation}
We have used a bar to denote the original fields and coordinates
$(\bar{x}, \bar{\Phi}, \bar{F}_{\mu\nu}^a, \ldots)$, we will
use the same letters without bar $(x, \Phi, F_{\mu\nu}^a,\ldots)$
for a rescaled version of these quantities (see below).

The non-Abelian field strenght tensor and the covariant derivative
of the Higgs field are given by
\begin{eqnarray}
  \bar{F}_{\mu\nu}^a & = & \bar{\partial}_\mu \bar{W}_\nu^a
         - \bar{\partial}_\nu \bar{W}_\mu^a
     + g \varepsilon_{abc} \bar{W}_\mu^b \bar{W}_\nu^c      \nn
  \mbox{and}\qquad
 \bar{{\cal D}}_\mu & = & \bar{\partial}_\mu
        - \frac{i}{2}g\tau^a \bar{W}^a_\mu
\end{eqnarray}
respectively, where $\tau^a$ are the Pauli matrices.

After spontaneous symmetry breaking the vacuum expectation
value of  $\bar{\Phi}$ is $v/\sqrt{2} $ and the
${W}_\mu^a$ bosons get a mass $ m_W = gv/2 $.
The mass of the physical Higgs boson becomes
$m_H=\sqrt{2 \lambda}\,v$. We define further, as usual,
the weak fine structure
constant as $ \alpha_w = g^2/4\pi$.

In view of the renormalization of the theory at finite temperature
it is suitable to rescale fields and coordinates as \footnote{Our
scale units differ from those of CLMW, we use $m_W$ instead of $gv$
as the basic unit.}
\beq
\label{xscale}
 \bar{x}_\mu = \frac{x_\mu}{m_W} = \frac{2x_\mu}{gv},
 \qquad \bar{W}_\mu^a = \frac{m_W}{g} W_\mu^a \qquad\mbox{and}\qquad
 \bar{\Phi} = \frac{v}{\sqrt{2}} \Phi.
\eeq

\noindent As a result of this rescaling the action takes the form
\beq
  \label{wirk}
  S = \frac{1}{g^2} \int d^4x \lag(x)
\eeq
with the Lagrangean density
\beq
  \label{lagdens}
  \lag = - \frac{1}{4} F_{\mu\nu}^a F^{\mu\nu a}
         + 2 \left({\cal D}_\mu\Phi\right)^\dagger
           \left({\cal D}^\mu\Phi\right)
         - \frac{1}{2}\xiq\left(\Phi^\dagger\Phi-1 \right)^2,
\eeq
Here we have introduced the ratio of Higgs and W masses $\xi$
via
\beq
  \label{xidef}
  \xiq \equiv \frac{m_H^2}{m_W^2} = \frac{8\lambda}{g^2}.
\eeq

Field strength and covariant derivative reduce to
\begin{eqnarray}
  F_{\mu\nu}^a & = & \partial_\mu W_\nu^a - \partial_\nu W_\mu^a
    + \varepsilon_{abc} W_\mu^b W_\nu^c                          \nn
  {\cal D}_\mu & = & \partial_\mu - \frac{i}{2} \tau^a W^a_\mu
\end{eqnarray}

As a consequence of this rescaling the action does not depend
any more on the vacuum expectation value of the Higgs field and
the couplings appear only in the mass ratio $\xi$.
As explained in  \cite{cmcl1,cmcl2} at high temperatures
the fields can be considered as essentially static and the
theory reduces to a three dimensional theory if the vacuum
expectation value is rescaled to the temperature dependent
one \cite{DoJa,ArMcL}
\bea
v(T) & = & v(0)\sqrt{1-T^2/T_c^2},                          \\
\mbox{where}\qquad
 T_c^2 & = & 2v(0)^2\left/\left[1+\frac{3g^2}{8\lambda}\right]
  \right.
\eea
is the critical temperature for symmetry restoration.
While the masses become then functions of the temperature,
$M_W(T)$ and $M_H(T)$, this dependence cancels
in their ratio $\xi$.

One obtains for the three dimensional Euclidean action
\beq
\label{euclidact}
  S_E = \frac{1}{g_3^2} \int d^3x \lag_E(x),
\eeq
with the Lagrangean density
\beq
  \lag_E(x) = \frac{1}{4} F_{\mu\nu}^a F_{\mu\nu a}
         + 2\left({\cal D}_\mu\Phi\right)^\dagger
           \left({\cal D}_\mu\Phi\right)
         + \frac{1}{2}\xiq\left(\Phi^\dagger\Phi-1 \right)^2.
\eeq
The effective coupling constant of the three dimensional theory
$g_3$ which describes the interaction at high temperatures
is given by
\beq
  g_3^2(T) \equiv \frac{g^2T}{M_W(T)}.
\eeq
For temperatures
\beq
  T \ll M_W(T)/\alpha_w
\eeq
this coupling is much smaller than $1$ and an expansion with respect
to $g_3$ should be reliable.

The variation of the action (\ref{wirk}) with respect
to $W_\mu^a$ and
 $\Phi^\dagger$ leads to the classical equations of motion
\begin{eqnarray}
  \label{cldgl1st}
  \left({\cal D}^\nu F_{\mu\nu}\right)^a
      + i\left[\left({\cal D}_\mu\Phi\right)^\dagger\tau^a\Phi
      - \Phi^\dagger\tau^a\left({\cal D}_\mu\Phi\right)\right]
      & = & 0                                                    \nn
  \mbox{and}\hspace{5cm}
  {\cal D}^\mu{\cal D}_\mu\Phi
      + \frac{1}{2}\xiq\left(\Phi^\dagger\Phi-1\right)\Phi
      & = & 0,
\end{eqnarray}
where the covariant divergence of the field strength tensor
is given by
\beq
  ({\cal D}^\nu F_{\mu\nu})^a = \partial^\nu F_{\mu\nu}^a
  + \varepsilon^{abc} W^{\nu b} F_{\mu\nu}^c
\eeq

In \cite{KlMa} a static saddle point solution to these equations
has been constructed explicitly, the well known sphaleron.
We choose here an Ansatz that differs from the one of
Ref. \cite{KlMa} by a $SU(2)$ rotation of the Higgs field
 by an angle
$\pi/2$ (which is another special case of a more general
parametrisation \cite{AkKiYa}) :
\begin{eqnarray}
  \label{clans}
  W_0^a & = & 0                                                 \nn
  W_j^a & = & \frac{f_A(r) - 1}{r}\varepsilon_{jam}\hat{x}_m   \nn
  \Phi & = &  H_0(r) {0 \choose 1}
\end{eqnarray}
with
\beq
  \hat{x}_m = \frac{x_m}{r}.
\eeq
Our solution will then be a gauge rotated version of the one given
in \cite{KlMa}.

Inserting this Ansatz into the classical equations of motion
(\ref{cldgl1st}) one obtains for the profile functions $f_A$
and $H_0$ a coupled system of the form
\begin{eqnarray}
  \label{cldgl}
  f''_A & = & (f_A-1)H_0^2 + \frac{f_A(f_A^2-1)}{r^2}           \nn
  H''_0 & = & \frac{1}{2}\xiq H_0(H_0^2-1) - \frac{2}{r}H'_0
              + \frac{(f_A - 1)^2}{2 r^2} H_0.
\end{eqnarray}

The energy of the sphaleron configuration
is given by
\beq
  \label{clenergy}
  E_{cl} = \frac{M_W(T)}{\alpha_w} \int dr\,{\cal H}(r),
\eeq
whith the Hamiltonian density
\beq
  \label{hamilton}
  {\cal H}(r) =  {f'}_A^2 + \frac{1}{2 r^2} (f_A^2 - 1)^2
        + 2 r^2 {H'}_0^2 +  (f_A - 1)^2 H_0^2
  + \frac{1}{2} \xiq r^2 (H_0^2 - 1)^2.
\eeq

Requiring the finiteness of the energy yields the following
boundary conditions for the profile
functions
\begin{eqnarray}
  \label{clrbd}
  f_A(0)      & = & -1                            \nn
  f_A(\infty) & = & 1                             \nn
  H_0(0)      & = & 0                             \nn
  H_0(\infty) & = & 1.
\end{eqnarray}

The numerical solutions of the equations (\ref{cldgl}) used later
on were obtained using a method developed in \cite{bp}.
The classical sphaleron energy as a function of
$\xi$ is presented in Fig.~1 in units of $ M_W(T)/\alpha_w$.
The classical Euclidean high temperature action as defined
in Eq. (\ref{euclidact}) is then given by
\beq
S^E_{cl}=E_{cl}/T.
\eeq

\setcounter{equation}{0}
\section{Small Fluctuations around the Sphaleron}
\label{qfspha}

In this section we will develop the small fluctuation expansion
around the sphaleron. We will start with expanding the Lagrangian
up to second order in the small fluctuations and fix the
gauge. Subsequently we will expand the small fluctuations
into partial waves.

The gauge and Higgs fields of the theory are expanded
around the spaleron configuration
via
\begin{eqnarray}
  W_0^a & = & a_0^a                                      \nn
  W_i^a & = & A_i^a + a_i^a
    \label{qfentw1}                                        \\
  \Phi   & = & (H_0+h) U(\varphi) {0 \choose 1}   \nn
         & = & \Pc + \Phi^{(1)}
  \label{qfentwphi}
\end{eqnarray}
\noindent
$A_i^a$ und $\Phi_{cl}$  denote here the classical solution
(\ref{clans})
\begin{eqnarray}
  \label{clans2}
  A_i^a & = & \frac{f_A - 1}{r} \varepsilon_{iam} \hat{x}_m   \nn
  \Phi_{cl} & = & H_0 {0 \choose 1}.
\end{eqnarray}
and $U(\phi)$ is given by
\beq
U(\varphi) = \exp ( i \tau^a \varphi^a).
\eeq
\noindent So the Higgs field fluctuations are parametrized by the
isosinglet $h$ and the isotriplet $\varphi^a$ and those of the gauge
field by $a_\mu^a$. We note that $\Phi$ has to be expanded up to
second order in the fluctuation fields if the second order
Lagrangean is to be determined; so $\Phi^{(1)}$ includes second
order terms \footnote{This has been done in exactly the same way in
\cite{BaLa}, but wrongly stated there in Eqs. (3.1) and (3.2)
which should be replaced by our Eqs. (\ref{qfentwphi})
and (\ref{qf_entw}).}.

Inserting these expansions into Eq. (\ref{lagdens}) and collecting
terms of the same order in the small fluctuations we find in zeroth
order the classical sphaleron action, the first order contribution
vanishes since the sphaleron configuration is a saddle point of the
action. The second order Lagrangean density in which we are
 interested here  becomes
\begin{eqnarray}
  \label{lag2}
  \lag^{(2)}
   & = & - \frac{1}{2}\left(\Dt_\mu a_\nu\right)^a
         \left(\Dt^\mu a^\nu\right)_a + \frac{1}{2}
         \left(\Dt_\mu a_\nu\right)^a
         \left(\Dt^\nu a^\mu\right)_a
         +\frac{1}{2}H_0^2 a_\mu^a a^{a \mu}  \nn
   &   & + 2 \partial_\mu h \partial^\mu h -\xi^2 \left(
   3 H_0^2-1 \right) h^2 + 2 \partial_\mu
   \varphi^a \partial^\mu \varphi^a  \nn
   &   & - \frac{1}{2}\eps^{abc} \Ft_{\mu\nu}^a a_b^\mu a_c^\nu
   -2 \eps^{abc} A_\mu^a \partial^\mu \varphi^b \varphi^c
    - 2H_0 a_\mu^a \partial^\mu \varphi^a  \nn
    &  & -4hA_\mu^a\partial^\mu \varphi^a + 2 H_0 h A_\mu^a a^{a\mu}
    + \frac{1}{2}A_\mu^a A^{a\mu} h^2
\end{eqnarray}
The tildes on the covariant derivatives indicate that only the
classical gauge fields are to be used:
\begin{eqnarray}
  \Ft_{\mu\nu}^a    & = &
     \partial_\mu A_\nu^a - \partial_\nu A_\mu^a
     + \varepsilon^{abc} A_\mu^b A_\nu^c          \nn
  \label{kova}
  (\Dt_\mu a_\nu)^a & = &
     \partial_\mu a_\nu^a + \eps^{abc} A_\mu^b a_\nu^c.
\end{eqnarray}
In order to eliminate the gauge degrees of freedom we have used the
background gauge which restricts the fluctuating fields.
In general form the three constraints are given by
\beq
  \label{zwang}
  {\cal F}_a = \left(\Dt_\mu a^\mu\right)^a
             - i\left[\Pc^\dagger\tau^a\Phi^{(1)}
             - \Phi^{(1)\dagger}\tau^a\Pc\right]
             =  0
\eeq
for $a=1,2,3$ where $\Phi^{(1)}$ is defined in Eq (\ref{qfentwphi}).
With our parametrization of the Higgs field they take the
form
\beq
\label{excons}
  {\cal F}^a = \left(\Dt_\mu a^\mu\right)^a + 2 H_0 \varphi^a.
\eeq
Here only the linear terms in $\Phi^{(1)}$ had to be taken into
account since only the square of ${\cal F}_a$ appears in the
gauge-fixed Lagrangean (see below).

 The Fadeev--Popov determinant for this gauge condition is
given by
\begin{eqnarray}
  \label{detfp}
    \det\Delta^{FP}_{ab} & = & \det\left[\Dt^2 +
     \Pc^\dagger\Pc\right]_{ab} \nn
                         & = & \det\left[\Dt^2 + H_0^2\right]_{ab}.
\end{eqnarray}

Finally one obtains the gauge fixed Lagrangean density
\begin{eqnarray}
  \label{lag_gf}
  \lag^{(2)}_{gf} & = & \lag^{(2)} - \frac{1}{2}
  {\cal F}^a{\cal F}^a  \nn
   & = & - \frac{1}{2}\left(\Dt_\mu a_0\right)^a
   \left(\Dt^\mu a_0\right)^a
         +\frac{1}{2} H_0^2 a_0^a a_0^a  \nn
   &  & + \frac{1}{2}\left(\Dt_\mu a_i\right)^a
                     \left(\Dt^\mu a_i\right)^a
         -\frac{1}{2}H_0^2 a_i^a a_i^a  \nn
   &   & + 2 \partial_\mu h \partial^\mu h -\xi^2 \left(
   3 H_0^2-1 \right)h^2  - \frac{1}{2} A_i^a A_i^a h^2 \nn
   &   & + 2 \partial_\mu \varphi^a \partial^\mu \varphi^a
   -2 H_0^2 \varphi^a \varphi^a  \nn
   &   & - \frac{1}{2}\eps^{abc} \Ft_{ij}^a a_i^b a_j^c
   +2 \eps^{abc} A_i^a \partial_i \varphi^b \varphi^c
    + 2 H_0 \eps^{abc } A_i^a a_i^b \varphi^c  \nn
    &  & +4hA_i^a\partial_i \varphi^a - 2 H_0 h A_i^a a_i^a
\end{eqnarray}

The background gauge has led to a decoupling of the
time components of the gauge fields. Their contribution
to the Lagrangean density
\begin{eqnarray}
  \lag^{(2)}_{a_0}
     & = & \frac{1}{2} a_0^a \left[\Dt^2
           +H_0^2\right]_{ab} a_0^b             \nn
     & = & \frac{1}{2} a_0^a \Delta^{FP}_{ab} a_0^b
\end{eqnarray}
leads to a fluctuation operator identical to that of the
Fadeev--Popov ghost fields.

\setcounter{equation}{0}
\section{Partial Wave Expansion of the Fluctuations}
In order to arrive at a suitable basis for our numerical
computations we have to decompose our system of small fluctuations
into partial waves with respect to the K spin $\vec K =\vec J +
\vec I$ combining angular momentum and isospin, which is conserved
on the sphaleron background. We avoid a large amount of
Clebsch--Gordan algebra by chosing a basis of tensor spherical
harmonics which is constructed using cartesian vector operators
(see e.g. \cite{HHSW}).

We define the following dimensionless operators
\beq
\label{opij_def}
\begin{array} {*{3}{r@{=}l}}
  \opj_a^1     &  \hat{x}_a                                    &
  \opj_a^2     &  r \nabla_a                                   &
  \opj_a^3     &  \eps_{abc} x_b \nabla_c \equiv \Lambda_a \\[0.5ex]
  \opi_{ja}^1  &  \hat{x}_j \hat{x}_a                          &
  \opi_{ja}^2  &  \hat{x}_j r \nabla_a                         &
  \opi_{ja}^3  &  r \nabla_j \hat{x}_a                     \\[0.5ex]
  \opi_{ja}^4  &  \hat{x}_j \Lambda_a                          &
  \opi_{ja}^5  &  \Lambda_j \hat{x}_a                          &
  \opi_{ja}^6  &  r \nabla_j r \nabla_a                    \\[0.5ex]
  \opi_{ja}^7  &  r \nabla_j \Lambda_a                         &
  \opi_{ja}^8  &  \Lambda_j r \nabla_a                         &
  \opi_{ja}^9  &  \Lambda_j \Lambda_a.
\end{array}
\eeq
These can be used to obtain a suitable basis of tensor spherical
harmonics with the spin--isospin transformation properties
appropriate to the different fields:
\begin{eqnarray}
  \label{qf_entw}
  a_{0a}    & = & \sum_{K,M} \sum_{\alpha=1}^3
                  s_\alpha(r) \opj_a^\alpha Y_{KM}(\hat{x})     \nn
  a_{ja}    & = & \sum_{K,M} \sum_{\alpha=1}^9
                  t_\alpha(r) \opi_{ja}^\alpha Y_{KM}(\hat{x})  \nn
  h         & = & \sum_{K,M} h_1(r)Y_{KM}(\hat{x}) {0 \choose 1}\nn
  \varphi_a & = & \sum_{K,M} \sum_{\alpha=1}^3 p_\alpha(r)
                  \opj_a^\alpha Y_{KM}(\hat{x}){0 \choose 1}.
\end{eqnarray}
The algebraic properties of the operators \opj ~and \opi ~and of the
tensor basis are discussed extensively in \cite{baamon} and will not
be repeated here. All manipulations have been performed using
REDUCE, starting with the basic Lagrangean density (\ref{lagdens})
and using the substitutions (\ref{qfentw1}), (\ref{qfentwphi}) and
(\ref{qf_entw}) and the gauge fixing based on (\ref{excons}).

The Lagrangean decomposes into contributions for each K spin.
 Varying these Lagrangeans one obtains a set of 16 linear coupled
 differential equations for the radial functions
$s_\alpha$, $t_\alpha$, $h_1$ and $p_\alpha$. According to
their parity transformation properties the 16 amplitudes
fall into two groups which do not couple between each other,
an electric  and a magnetic sector. The electric sector consists
of the amplitudes $t_4$, $t_5$, $t_7$, $t_8$, $p_3$, $h_1$ and
$s_3$ with parity $\pi=(-1)^{K+1}$ and the magnetic sector contains
the amplitudes $t_1$, $t_2$, $t_3$, $t_6$, $t_9$, $p_1$, $p_2$,
$s_1$ and $s_2$ with parity $\pi=(-1)^K$. Furthermore the ghost
amplitudes $s_1$, $s_2$ and $s_3$ are not coupled to the other
fields. They form the Fadeev--Popov sector. So altogether we have a
$6 \times 6$ and a $7 \times 7$ subsystem for the gauge and Higgs
fields. Only $s_1$ and $s_2$ form a coupled system in the
Fadeev--Popov sector but we will not divide this sector furthermore.
We will refer to the three sectors by a symbol $\sigma$ which
can have the values $E, M$ and $FP$  and the different
coupled channnels will be referred to by the two quantities $\sigma$
for the fields and $K$ for the partial wave.

The K spins  $K=0$ and $K=1$ have to be considered separately:
In the case $K=0$ the \opj-- and \opi-- operators act on constant
spherical harmonic $Y_{00}$ so that only the operators $\opi^1$,
$\opi^3$, $\opi^5$ and $\opj^1$ contribute; the amplitudes $t_2$,
$t_4$, $t_6$, $t_7$, $t_8$, $t_9$, $p_2$, $p_3$, $s_2$ and $s_3$
have to be discarded and we have only 6 fluctuation equations.
In the case $K=1$ the action of the operators $\opi^3$ and $\opi^6$
on the spherical harmonics $Y_{1M} \propto \hat{x}$ yields identical
tensors and the same is true for $\opi^5$ and $\opi^8$:
\begin{eqnarray}
  \label{dsfa}
    \opi^3_{ia} \hat{x}_j = - \opi^6_{ia} \hat{x}_j & = &
        \delta_{ia}\hat{x}_j+\delta_{ij}\hat{x}_a
        -2\hat{x}_i\hat{x}_a\hat{x}_j                           \nn
    \opi^5_{ia} \hat{x}_j = - \opi^8_{ia} \hat{x}_j & = &
      \left(\eps_{ika}\hat{x}_j+\eps_{ikj}\hat{x}_a\right)\hat{x}_k.
\end{eqnarray}

One of the amplitudes $t_3$ and $t_6$ and one of the amplitudes
$t_5$ und $t_8$ have to be discarded. We have chosen
\beq
  t_6 = t_8 = 0, \qquad\mbox{for}\;K=1.
\eeq
This elimination cannot be done simply in the general fluctuation
equation but has to be performed already in the Lagrangean.

For the following general developments we will design by
$\bar{\Psi}$ a column vector which is formed by the amplitudes of
one of the coupled sectors. The fluctuation equations can then be
written in the general form
\beq
  \label{flukdgl1}
  - \ddot{\bar{\Psi}}_i + \bar{\Psi}''_i + \frac{2}{r} \bar{\Psi}'_i
       - m_i^2\bar{\Psi_i} = \bar{V}_{ij}\bar{\Psi}_j.
\eeq
They are given explicitly in Appendix \ref{eqmo}. As one can see
there the fluctuation equations have some shortcomings:
the amplitudes do not display a unique centrifugal term, the
potential $\bar{V}_{ij}$ is not symmetric and in the vacuum, i.e.
for $(f_A \to 1, H_0 \to 1, f'_A \to 0, H'_0 \to 0)$ the amplitudes
are still coupled. One has to find therefore a transformation
\beq
  \bar{\Psi}_i = C_{ij} \Psi_j
\eeq
of the fields that brings them to a form
\beq
  \label{flukdgl2}
  - \ddot{\Psi}_i + \Psi''_i + \frac{2}{r} \Psi'_i - m_i^2\Psi_i
  - \frac{l_i(l_i + 1)}{r^2}\Psi_i = V_{ij}\Psi_j
\eeq
where the potential satisfies
\beq
  V_{ij}=V_{ji} \qquad\mbox{und}\qquad \lim_{r\to\infty} V_{ij} = 0.
\eeq
Such a transform must obviously exist since the Lagrangean
(\ref{lag_gf}) is a symmetric bilinear form in the fields.

The manipulations leading to this transform have been
performed using REDUCE. The new amplitudes, the associated
angular momenta and masses and the potential in the new basis
are also given in Appendix \ref{eqmo}. The equations in this
form constitute the
basis of our further developments.
\setcounter{equation}{0}
\section{The Fluctuation Determinant ${\bf\kappa}$}
\label{green}

The rate for sphaleron transitions -- based on the general
theory of Langer -- has been obtained by various authors
(\cite{Af,ArMcL,AkKiYa}) to be given by
\beq
\Gamma /V = \frac{\omega_-}{2 \pi} T^{-3} {\cal N}_{tr}
{\cal N}_{rot} V_{rot} \exp(-E_{cl}/T) \kappa.
\eeq
Here $\kappa$ and the other prefactors arise from the Gaussian
functional integration of the quadratic fluctuations around the
classical saddle point solution. $\kappa$ is essentially (see below)
the determinant corresponding to the fluctuation operator
\beq
  \label{flukdel}
    \Delta_{ij} = \left.\frac{\delta^2S_{gf}^{(2)}(\psi)}
      {\delta\psi_i\,\delta\psi_j}\right|_{\psi=\Psi_{cl}}.
\eeq
The calculation of $\kappa$ is, besides the general partial wave
analysis, the main subject of this publication.

The prefactors labelled $tr$ and $rot$ are related to the
existence of six zero modes due to the invariance of the theory
with respect to translations and rotations which
is broken by the sphaleron solution. These modes satisfy
\beq
  \Delta_{ij}\Psi^{rot}_j = \Delta_{ij}\Psi^{tr}_j = 0.
\eeq
The functional integration of these modes has to be treated
separately (see  \cite{La,ArMcL,cmcl1,cmcl2}). They have to be
excluded in evaluating the fluctuation determinant.
These prefactors and their calculation are discussed in Appendix B.
The results are displayed in Fig. 2. They agree within a few percent
with those of CLMW apart from factors $2^{\pm 3/2}$ arising
from the different scale factors $M_W$ resp. $gv$ used to make
the radial variable dimensionless (see Eq. (\ref{xscale})).

As obvious from the discussion in the previous section K spin and
parity invariance lead to a decomposition of the fluctuation
operator into a direct sum of fluctuation operators within the
different coupled channels $\Delta_{\sigma K}$. On account of
Eq. (\ref{flukdgl2}) these operators take the form
\beq
  \label{deldef}
  \Delta_{(\sigma K)ij} \equiv \left(\frac{d^2}{dr^2}+\frac{2}{r}\,
  \frac{d}{dr}-m_i^2-\frac{l_i(l_i+1)}{r^2}\right)\delta_{ij}
  -V_{(\sigma K)ij}.
\eeq
The angular momenta $l_i$, masses $m_i$ and potentials
$V_{(\sigma K) ij}$ of these operators in the various channels are
given in Appendix A. The fluctuation determinant decomposes
accordingly. Therefore
\beq
  \label{lcalk}
  \ln \kappa
     = \frac{1}{2}\sum_{K=0}^\infty (2K+1)
       \left[\ln\frac{\det\Delta^{(0)}_{M K}}{\det'\Delta_{M K}}
       +\ln\frac{\det\Delta^{(0)}_{E K}}{\det'\Delta_{E K}}
       -\ln\frac{\det\Delta^{(0)}_{FP K}}{\det\Delta_{FP K}}\right].
\eeq
Here it has been used that the sector consisting of the time
components of the gauge field has the same fluctuation operator as
the Fadeev--Popov sector. The minus sign in front of the
Fadeev--Popov contribution arises from a factor $(1-2)$ where the
$1$ correspond to the time components of the (real) gauge field and
the $-2$ to the (complex) Fadeev--Popov ghosts. The apostrophes $'$
indicate that in the evaluation of the determinants the zero modes
have to be removed and the negative squared frequency of the
unstable mode has to be replaced by its absolute value.

\setcounter{equation}{0}

\section{The Numerical Procedure}
\label{num_met}

Consider the contribution of one of the coupled channels
$(\sigma K)$ to the logarithm of the fluctuation determinant
$\kappa$
\beq
  \label{kbeitrag}
 (\ln \kappa)_{\sigma K} \equiv
 \frac{1}{2}\ln\frac{\det\Delta^{(0)}_{\sigma K}}
 {\det'\Delta_{\sigma K}}
\eeq
with the operators  $\Delta_{\sigma K}$ and
$\Delta^{(0)}_{\sigma K}$ as defined in Eq. (\ref{deldef}).
Since our discussion will be confined just to one channel the
indices $K$ and $\sigma$ are omitted in the following from the
operators $\Delta$, their Green functions, eigenfunctions and
eigenvalues.

Let us define the Green function $G_{ij}(r,r',\nu)$ by
\beq
  \label{greendef}
  \left[\Delta_{ij}-\nu^2\delta_{ij}\right]
         G_{jk}(r,r',\nu)
      = -\frac{1}{r^2}\delta(r-r')\delta_{ik}
\eeq
and the analoguous Green function $G^{(0)}$ for the
operator $\Delta^{(0)}$. The solution to this equation can be
written (using discrete notation formally) as
\beq
  \label{greenfun}
  G_{jk}(r,r',\nu)
      = \sum_\alpha \frac{\phi_{j\alpha}(r)\phi^\ast_{k \alpha}(r')}
      {\lambda^2_{\alpha}+\nu^2}
\eeq
in terms of the orthonormalized eigenfunctions  $\phi_{j \alpha}$
of the fluctuation operator $\Delta$. The latin letters refer to
the field components and the greek letters label the different
eigenmodes.  Analoguous relations hold for the vacuum Green
function. We define now a function $F_{\sigma K}(\nu)$ by
\beq
  F_{\sigma K}(\nu)=\int dr r^2 (G_{ii}(r,r,\nu)-G^{(0)}_{ii}
  (r,r,\nu)).
\eeq
Using the expression (\ref{greenfun}) it can be written
in terms of the eigenfrequencies $\lambda_\alpha$ as
\beq
F_{\sigma K}(\nu)= \sum_{\alpha}( \frac{1}{\lambda^2_{\alpha}+\nu^2}
-\frac{1}{{\lambda^{(0)}}^2_{\alpha}+\nu^2}).
\eeq

Integrating this expression over $\nu\,d\nu$ from $\epsilon$ to
$\Lambda$ yields
\beq
\label{PaVidef}
\int_\epsilon^\Lambda d\nu\, \nu F_{\sigma K}(\nu)=
\frac{1}{2}\sum_\alpha(\ln(\frac{{\lambda^{(0)}}^2_\alpha+\epsilon^2
}{\lambda^2_\alpha+\epsilon^2})
-\ln(\frac{{\lambda^{(0)}}^2_\alpha+\Lambda^2}{\lambda^2_\alpha
+\Lambda^2})).
\eeq

Taking into account the fact that the operators $\Delta$ and
$\Delta^{(0)}$ and therefore their eigenva\-lues $\lambda^2$
are linear in the $m_i^2$ this expression is a
Pauli--Villars regulated version of $(\ln \kappa)_{\sigma K}$, the
second term in the sum being obtained by replacing all $m_i^2$ by
$m_i^2+\Lambda^2$.
Of course we should let $\Lambda \to \infty$ after the
expressions have been renormalized and the integral has become
ultraviolet convergent. Renormalization will be discussed in section
\ref{renormalization}.

The lower limit $\epsilon$ should of course be set equal to zero.
We have introduced it because for $K=1$ we have six zero modes and
the limit $\epsilon \to 0$ then obviously does not exist. We will
deal with this problem below (see section \ref{zeromod})
when we discuss the removal of these modes.

A further problem arises from the unstable mode which leads to
a pole in the region of integration and requires a precise
definition of the integration contour. This problem will be tackled
also in section \ref{zeromod}.

Leaving aside these problems for the time being we have
the ``naive'' relation
\beq
  \label{lnk2}
(\ln \kappa)_{\sigma K} = \int_0^\infty d\nu\,\nu F_{\sigma K}(\nu).
\eeq

In order to evaluate $F_{\sigma K}(\nu)$ the Green functions have
to be determined. We will not use their expansion with
respect to eigenfunctions (see Eq. (\ref{greenfun})), but use
another standard method:

Let ${f_n^\alpha}^\pm$ be the solutions of the homogeneous
 differential equations
\beq
  \label{fdgl}
  \left(\frac{d^2}{dr^2} + \frac{2}{r}\frac{d}{dr}
  - \frac{l_n(l_n + 1)}{r^2} - \kappa_n^2\right) {f_n^\alpha}^\pm(r)
  = V_{nn'}(r) {f_{n'}^\alpha}^\pm(r),
\eeq
where  $\kappa_n$ has been defined as\footnote{The letter $\kappa$
has been introduced previously to denote the fluctuation
determinant. Since $\kappa_n$ as defined here will always
appear with an index there should be no confusion.}
\beq
\label{kapdef}
  \kappa_n \equiv \sqrt{\nu^2 + m_n^2}
\eeq
whith $m_n = 1$ or $\xi$ depending on the field component $n$.
$f_n^{\alpha+}$ designs the solution regular as $r\to \infty$
and $f_n^{\alpha-}$ the solution regular as $r\to 0$. The index
$n$ labels again the field components, the greek letters will in the
following label a set of linearly independent solutions of the
system; there are of course as many such solutions as there are
field components.

For the vacuum ($V_{nn'}=0$) these equations are solved
by modified Bessel functions with the argument
$z_n=\kappa_n r$, which are defined here (slightly deviating
from \cite{abra} ) as
\begin{eqnarray}
  i_l(z) & = & \sqrt{\frac{\pi}{2z}} I_{l+\frac{1}{2}}(z)    \nn
  k_l(z) & = & \sqrt{\frac{2}{\pi z}} K_{l+\frac{1}{2}}(z).
\end{eqnarray}

Their Wronskian is given by
\beq
  W(i_{l_n},k_{l_n})
     = \kappa_n \left(i_{l_n}(z_n)k'_{l_n}(z_n)
           - k_{l_n}(z_n)i'_{l_n}(z_n)\right)
     = - \frac{1}{\kappa_n r^2}.
\eeq

The behaviour of the solutions ${f_n^\alpha}^\pm$ for $r\to 0$
and $r\to\infty$ is analogous to the one of these free solutions:
\beq
  \left.{i_l \propto z^l \atop k_l \propto z^{-(l+1)}}\right\}
      \mbox{for}\; z\to 0
  \qquad\mbox{and}\qquad
  \left.{i_l \propto e^z/z \atop k_l \propto e^{-z}/z}\right\}
      \mbox{for}\; z\to\infty.
\eeq

It is convenient to split off the Bessel functions from
the solutions $f_n^{\alpha \pm}$ via
\beq
  \label{glfh}
  f_n^{\alpha \pm}(r) = \left[\delta_n^\alpha
      + h_n^{\alpha\pm}(r)\right] {b_{l_n}}^\pm(z_n)
\eeq
with
\beq
   {b_l}^- = i_l  \qquad\mbox{and}\qquad {b_l}^+ = k_l.
\eeq

On account of the boundary conditions for the functions
${f_n^\alpha}^\pm$ the functions ${h_n^\alpha}^\pm$
tend to a constant as $r\to\infty$. The choice
\beq
  \label{hrbd}
  \lim_{r\to\infty} h_n^{\alpha\pm}(r) =
  \lim_{r\to\infty} {h'}_n^{\alpha\pm}(r) = 0
\eeq
determines the normalization of the amplitudes $f_n^{\alpha\pm}$
in such a way that their Wronskian satisfies
\beq
  \label{wronf}
  W^{\alpha\beta}(r)
      =\sum_n \left({f_n^\alpha}^+\frac{d}{dr}{f_n^\beta}^-
            -{f_n^\beta}^-\frac{d}{dr}{f_n^\alpha}^+\right)
      = \frac{1}{\kappa_\alpha r^2} \delta^{\alpha\beta}.
\eeq
For $r\to 0$ the centrifugal barriers for the $f_n^{\alpha \pm}$
 are different from the
ones for the free solutions. Effectively they differ at most by two
units from those, so that the functions $\delta^\alpha_n+h_n^
{\alpha\pm}$ behave as $r^0,r^{\pm 1}$   or $r^{\pm 2}$
in this limit; most of the typical $r^{l_n}$ behaviour that is
dangerous for numerical calculations in high partial waves
has however been taken out by splitting off the free solutions.

For our new functions $h_n^{\alpha \pm}$ equations (\ref{fdgl})
yield the inhomogeneous differential equations
\beq
  \label{hdgl}
  \left[\frac{d^2}{dr^2} + 2 \left(\frac{1}{r} + \kappa_n
  \frac{b'_{l_n}(z_n)}{b_{l_n}(z_n)} \frac{d}{dr} \right)\right]
  h_n^{\alpha \pm}(r) = V_{nn'} \left[\delta_{n'}^\alpha
  + h_{n'}^{\alpha \pm}(r)\right] \frac{b_{l_{n'}}(z_{n'})}{b_{l_n}
  (z_n)}.
\eeq
They can be integrated numerically using the
Nystr\"{o}m  method (Runge  Kutta  integration for second order
differential equations \cite{nyst}).
Some numerical details are discussed in Appendix C.

{}From these solutions the Green function defined in
Eq. (\ref{greendef}) is now obtained as
\beq
  \label{greenexakt}
  G_{nn'}(r,r',\nu) = \Theta(r-r'){f_n^\alpha}^+(r)
      C_{\beta\alpha}^{-1} {f_{n'}^\beta}^-(r') + \Theta(r'-r)
      {f_n^\alpha}^-(r) C_{\alpha\beta}^{-1}
      {f_{n'}^\beta}^+(r')
\eeq
where the coefficients
$C_{\alpha\beta}$ are related to the Wronskian of the
amplitudes ${f_n^\alpha}^\pm$ as
\beq
  C_{\alpha\beta} = r^2 W_{\alpha\beta}(r)
                  = \frac{1}{\kappa_\alpha}\delta_{\alpha\beta}.
\eeq

With these preliminaries we obtain for the trace of the
Green functions at  $r'=r$:
\begin{eqnarray}
  G^{(0)}_{nn}(r,r,\nu)
     & = & \sum_n \kappa_n i_{l_n}(z_n) k_{l_n}(z_n)          \nn
  G_{nn}(r,r,\nu)
     & = &\sum_{\alpha,n}
      \kappa_\alpha{f_n^\alpha}^-(r){f_n^\alpha}^+(r)
\end{eqnarray}

Therefore, using Eq. (\ref{glfh}), the
function $F_{\sigma K}(\nu)$ can be calculated for each
system of coupled channels $(\sigma K)$ as
\beq
\label{Feval}
  F_{\sigma K}(\nu) = \int_0^\infty\,dr\,r^2 \sum_{\alpha,n}
      \kappa_\alpha \left[\delta_n^\alpha\left(h_n^{\alpha-}
        + h_n^{\alpha+} \right) + h_n^{\alpha-} h_n^{\alpha+}
        \right]i_{l_n}k_{l_n}
\eeq
once the functions $h_n^{\alpha \pm}$ for that channel
have been found numerically.

The total fluctuation determinant is then obtained as
\beq
\label{lnkk1}
\ln \kappa = \int_0^\infty d\nu\,\nu\, F(\nu)
\eeq with
\beq
\label{lnkk2}
F(\nu)=\sum_{K=0}^\infty (2K+1) (F_{E K}(\nu)
+F_{M K}(\nu)-F_{FP K}(\nu))
\eeq

This expression is still formal, we have to discuss the treatment
of zero and unstable modes and of renormalization.
\setcounter{equation}{0}
\section{Renormalization}
\label{renormalization}
As has been discussed by Carson et al. \cite{cmcl2}
renormalization requires, in the $T \to \infty$ limit discussed
here, the replacement of the Higgs vacuum expectation value by its
temperature dependent value and the subtraction of the tadpole
graphs with Higgs fields as external legs.

In order to do this we discuss first the
relation of our expression for $\ln \kappa$ of Eqs.
(\ref{Feval})--(\ref{lnkk2})
to a Feynman graph expansion. The function $F(\nu)$ is obtained
as the trace of the Green function of the ``small fields''
$a_i$, $\eta$ and the Fadeev--Popov fields in the external
potential generated by the classical fields. It has been
obtained here by decomposing
this Green function first into partial waves and then summing
over the individual partial wave contributions.
This Green function may also be expanded with respect to the
external potential; formally
\beq
\label{Greenexp}
G_{ij}(\vec x,   \vec x', \nu)=
<\vec x ,i|\sum_{n=0}^\infty [\frac {-1}{{\bf p}^2+{\bf m}^2+\nu^2}
{\cal \bf V}({\bf x})]^n \frac{1}{{\bf p}^2+{\bf m}^2+\nu^2}
|\vec x',j>
\eeq
where bold face letters indicate operators and/or matrices.
Then
\beq
F(\nu)=\int d^3 x G_{ii}(\vec c, \vec x, \nu)
\eeq
and therefore the fluctuation determinant is obtained via
\bea
\ln \kappa &=& \int d\nu\,\nu\,F(\nu)   \nn
     &=& \sum_{n=1}^\infty \frac {1}{2n}\int d^3x <\vec x,i|
    [ \frac {-1}{{\bf p}^2+{\bf m}^2}{\cal \bf V}({\bf x})]^n
    |\vec x,i>
 \eea
which is just the normal Feynman graph expansion of the
1--loop effective action for a 3 dimensional theory.

The tadpole contributions to $F(\nu)$ are therefore obtained as
that part of the first order contribution in the external
potential that is generated by the classical Higgs field.
These terms are easily recognized in the
second order Lagrangian ${\cal L}_{gf}^{(2)}$ presented
in Eq. (\ref{lag_gf}) as those containing
$\Phi_{cl}^2$, $\Phi_{cl}^{\dagger 2}$ and $\Phi_{cl}^\dagger\
Phi_{cl}$. In the partial wave potential given explicitly in
Appendix A these terms appear in the diagonal elements and
are proportional to $(H_0^2-1)$.

The tadpole terms can therefore be subtracted either in the
single partial waves as the first order perturbative
contribution generated be the $(H_0^2-1)$ terms or
directly from $F(\nu)$.

For each partial wave the differential equation for the
radial wave function may be transformed into an integral
equation and this integral equation can be used
for a pertur\-bative expansion; this has been discussed extensively
in \cite{baanum}. The first order contribution of such an
expansion is obtained as
\beq
  \label{lnk1b}
  (\ln \kappa)_{\sigma K}^{(1)}
       = -  \int_0^\infty d\nu\,\nu\,\kappa_n^2
      \int_0^\infty\,dr\,r^2
       \int_0^\infty dr'\,{r'}^2 \tilde{V}_{nn}(r')
            i_{l_n}^2(\kappa_n r'_<)k_{l_n}^2(\kappa_n r'_>)
\eeq
where $r'_< \equiv \min(r,r')$ and $r'_> \equiv \max(r,r')$.
The tilde over the potential indicates that only the terms
proportional to $(H_0^2-1)$ are to be included.

One of the radial integrations can be performed
analytically so that
\beq
  \label{lnk1}
  (\ln \kappa)_{\sigma K}^{(1)}
    =  \frac{1}{2} \int_0^\infty d\nu\,\nu\int_0^\infty dr\,r^3
   \tilde{V}_{nn}(r)\left[i_{l_n-1}k_{l_n}-k_{l_n+1}i_{l_n}\right].
\eeq
Either by summing up the partial wave contributions or by
calculation of the Feynman type graphs according to Eq.
(\ref{Greenexp}) the total tadpole contribution to $F(\nu) $ is
obtained as
\beq
\label{Ftad}
F_{tad}(\nu) =-\frac{1}{2} \int_0^\infty dr\,r^2 (H_0^2(r)-1)
[\frac{9}{\kappa_W}+ \frac{3\xi^2}{2\kappa_H}+\frac{3\xi^2}
{2\kappa_W}]
\eeq
where $\kappa_n$ has been defined in Eq. (\ref{kapdef}).
The renormalized value of $\ln \kappa$ is therefore obtained from
\beq
F_{ren}(\nu)=F(\nu)-F_{tad}(\nu)
\eeq
where the subtraction can be done either in the partial waves using
Eq. (\ref{lnk1}) or in the full amplitude using Eq. (\ref{Ftad}).
The $\nu$ integration  of $F_{ren}(\nu)$ is now ultraviolet
convergent, i.e. the upper limit of integration $\Lambda$ introduced
in Eq. (\ref{PaVidef}) which serves also as a Pauli-Villars
regulator can be sent to $\infty$.

\setcounter{equation}{0}
\section{Zero and Unstable Modes}
\label{zeromod}
Besides the question of renormalization we have also postponed
the treatment of the zero and unstable modes.

The zero modes should be removed from the fluctuation determinant.
Equation (\ref{PaVidef}) shows that their contribution to
$(\ln \kappa)_{E1}$ and $(\ln \kappa)_{M1}$ is in both cases
$-(1/2) \ln(\lambda_0^2+\epsilon^2)$ before $\epsilon$ goes
to $0$. Of course $\lambda_0 =0$ here and the limit does not exist.
But we have to remove just these contributions. Therefore the
fluctuation determinant without the zero modes is obtained as
\beq
\label{remzermod}
\ln \kappa = \lim_{\epsilon \to 0}
(\int_\epsilon^\infty d\nu\,\nu F_{ren}(\nu) + 6 \ln (\epsilon)).
\eeq

The function $F_{ren}(\nu)$ will of course
behave as $6/\nu^2$
due to the zero modes so that the limit exists.
It has to do so also in the numerical evaluation; this represents
a good cross check.
Of course $\epsilon$ is a quantity of dimension energy and
indeed removing the zero modes makes $\kappa$ a quantity of
dimension $(energy)^6$. This has to be taken into account when
comparing results if different length and energy units are used.
We have used units of $M_W^{-1}$ for the radial variable;
our unit for the eigenvalues of the second order differential
operators $\Delta$ is therefore $M_W^2$ ,their eigenfrequencies
and therefore also the variables $\nu$ and $\epsilon$ are
therefore in units $M_W$.

The unstable mode is not to be removed but, according to
the general theory \cite{La,Af} to be replaced by its
absolute value. If its eigenvalue is denoted
by $\lambda_-^2=-\omega_-^2 < 0$ then it leads to a pole
in $F(\nu)$ at $\nu^2=\omega_-^2 >0$ and therefore in the region
of integration of Eq. (\ref{lnkk1}).
It would contribute a term $-(1/2) \ln (-\omega_-^2)$
and the minus sign in the logarithm
should be removed. This can be done simply
by evaluating the integral as a principal value integral.

Since doing singular principal value integrals numerically
is a delicate operation we have subtracted the pole from the
integrand and done its integration analytically.
We have used the identity
\bea
  \label{instmod}
  (\ln\kappa)& = & P \int_0^\infty d\nu\,\nu\,F(\nu)         \nn
  &=&     \int_0^\infty d\nu\,\nu\left[F_{ren}(\nu)
            - \frac{1}{\nu^2-\omega_-^2}+\frac{1}{\nu^2+\sigma^2}
            \right] - \ln(\omega_-/\sigma).
\eea
The third term in the parenthesis was added in order not to
spoil the ultraviolet convergence of the integral.
Since (\ref{instmod}) is an identity the value of
$\sigma$ is in principle arbitrary. We have chosen
$ \sigma =1 $
in units of $M_W$ for convenience.

For easier notation have treated the removal of the zero modes and
of the unstable mode singularity separately. It is of course
understood that both prescriptions Eqs. (\ref{remzermod}) and
(\ref{instmod}) are applied simultaneously.

\setcounter{equation}{0}
\section{Results}
\label{results}

We have now discussed the principles of our numerical
procedure for calculating the fluctuation determinant.
Before presenting the results we will discuss some specific
details of our numerical evaluation.

The first step was the Runge--Kutta integration of the partial wave
differential equations (\ref{hdgl}) in each channel
and the numerical integration of the  exact trace
as presented in  Eqs. (\ref{Feval})--(\ref{lnkk2}). The numerical
details are discussed in Appendix C.

In order to obtain the function $F(\nu)$ we had to sum over all
partial waves $K$ within the various sectors $\sigma$ and then to
perform the summation over $\sigma$. In order to have a check on the
$K$ summation we have considered the asymptotic behaviour
at large $K$ of the terms in this sum. For this purpose it
is sufficient to consider the perturbative contribution
of first order in the potentials $V_{ij}$, higher order
contributions will decrease faster. The first order
contributions to the sum in Eq (\ref{lnkk2}) have the form
\beq
(2K+1) F_{\sigma K}^{(1)}=
(2K+1)\int_0^\infty dr \frac{r^3}{2} V_{nn}(r)\left[i_{l_n-1}k_{l_n}
-k_{l_n+1}i_{l_n}\right].
\eeq
Using the uniform asymptotic expansion of the Bessel
functions at large $K$ \cite{abra}
one finds that this expression behaves as $K^{-2}$ asymptotically.
This determines then the convergence behaviour of the $K$ summation.
It has been checked numerically to a good accuracy. Since we know
the leading behaviour, we can extrapolate the terms to arbitrary
valus of $K$. We have done so by fitting the last five calculated
values with a power behaviour
\beq
\label{asfit}
(2K+1) F_{\sigma K}^{(1)} \simeq \frac{C_2}{(2K+1)^2}+\frac{C_3}
{(2K+1)^3}
\eeq
To the sum extended to some value $K_{max}$
\beq
F(K_{max},\nu)=\sum_{K=1}^{K_{max}} \sum_{\sigma}
(2K+1) F_{\sigma K}
\eeq
we have then added the sum from $(K_{max}+1)$ to $\infty$ using the
fit (\ref{asfit}).
Fig. 3 shows the partial sums as well as the sums completed
using the extrapolated values. One sees that the complete sum
becomes independent of $K_{max}$ already at moderate values of
this variable. This shows that the extrapolation procedure is
reliable.

The function $F(\nu)$ is displayed in Fig. 4 for $\nu = 1$.
The pole contribution of the unstable mode is already
subtracted here according to Eq. (\ref{instmod}).
The dashed line shows the full function $F(\nu)$, the
dash--dotted line the tadpole contribution. Note that this
contribution is determined analytically in absolute
normalization. So the fact that both curves approach each
other as $\nu \to \infty$ checks also the absolute normalization
of the unsubtracted $F(\nu)$. The full line shows $F_{ren}(\nu)$
together with the asymptotic estimates at small and large $\nu$
(dotted lines). The behaviour at small $\nu$ is normalized
absolutely; it is determined by the zero modes to be
$6/\nu^2$.

The $\nu$ integration was performed numerically up to
$\nu_{max} \simeq 2.5$, then an asymptotic part was added to the
integral by extrapolating $F_{ren}$ as
$C_i/\kappa_i^3+D_i/\kappa_i^5$. The
results for $\ln \kappa$ are given in Table \ref{restab} for various
values of $\xi=M_H/M_W$ in the scale $M_W$. They are plotted in
Fig. 5 together with previous results and estimates in the scale
$gv$, i.e. after subtracting $6\ln 2$ from the values of Table
\ref{restab}. Our calculation stops at $M_H/M_W = 2$ for a technical
 reason: Above this value the leading
asymptotic behaviour $\exp (-\kappa_H r)$ of the fluctuation $h_1$
of the Higgs field  becomes dominated -- through a cross term
in the potential -- by a gauge
field contribution which behaves as $\exp(-(M_W+\kappa_W)r)$. So
for $M_H >2 M_W$ the boundary conditions for this function
have to be modified.
If the Higgs mass should turn out to be larger than this value
 we would have to deal with this complication.

We think that the main uncertainty of our results comes from
the extrapolation of $F_{ren}(\nu)$ to $\nu=\infty$. We estimate the
error of this asymptotic contribution to be around 10\% yielding an
typical error of $0.3$ for $\ln \kappa$ and we think that this is a
conservative estimate.

Our results differ considerably from the ones of
CLMW \cite{cmcl2}. There is one point which could lead to a
difference on physical grounds: we have used a different gauge
in the classical sphaleron ansatz (see below Eq. (\ref{clans})).
While the sphaleron transition rate must be gauge invariant,
a difference in the classical configuration could modify the zero
mode prefactors and then require also a compensating change in the
fluctuation determinant \footnote{We thank L. McLerran for pointing
this out to us.}. However -- as mentioned above --
the prefactors agree and therefore should also the fluctuation
determinants.

The methods used by CLMW and by us differ also considerably.
Since we work with the Euclidean Green function, we have no
difficulties with the fact that the spectrum of the various
fluctuation operators is continuous (except for zero and
unstable modes and a few further bound states). The Schwinger
proper time method as used by CLMW requires a discrete spectrum
which has to be created artificially by introducing a
space boundary (a kind of ``bag'') of radius $R$. The calculation
requires then a limit $R \to \infty$. This limiting procedure
is absent in our method. This makes the algorithm much faster.
Furthermore the method contains more internal consistency
checks. Of course this does not necessarily imply that our
results are the correct ones but at present we have not found
any reason to doubt them. We hope that the discrepancy
can be settled in the near future, possibly by another
calculation.

Comparing to the DPY approximation we find that our results come,
at $M_H > M_W$, much closer to this approximation than the ones of
CLMW; they show a different trend for small Higgs masses, however.
The validity of the DPY approximation  has been discussed by
Carson \cite{Ca} for a one--dimensional sphaleron where the exact
fluctuation determinant is known analytically. He finds only
fair agreement between the exact results and the approximation.
Furthermore (see \cite{BaSche}) it is not obvious in which way
the convergence of gradient expansions (and the DPY approximation
falls under this category) in $1+1$ and  $3+1$
dimensions can be compared.

In conclusion we have presented here a new set of fluctuation
equations for the the electroweak sphaleron at $\Theta_W=0$. We have
used it to evaluate the fluctuation determinant. We obtain the
 result that the quantum corrections lead to an enhancement of the
sphaleron transition with respect to the estimate obtained
from the classical saddle point solution.
\newpage
\begin{appendix}
\setcounter{equation}{0}
\section{The Partial Wave Fluctuation Equations}
\label{eqmo}

The equations of motion have been obtained as follows:
the Ansatz for the fluctuation amplitudes was inserted into
the second order fluctuation Lagrangean which itself had
been obtained using REDUCE from the general gauge-fixed
Lagrangean. A second REDUCE step used the
various algebraic properties of the operators \opi~and
\opj~given in \cite{baamon} in order to obtain a Lagrangean
quadratic in the amplitudes $t_\alpha$, $s_\alpha$, $p_\alpha$
and $h$. The Lagrangean before gauge fixing had already be
obtained in \cite{BaLa}. Checking its gauge invariance
and the presence of all zero modes presents a good test
on its correctness.

The equations of motion obtained in this way are -- for
$K > 1$ -- given by
\begin{eqnarray}
r^2t''_1 & + & 2rt'_1 - r^2\ddot{t}_1                          \nn
  & = & t_1(r^2H_0^2 + 2f_A^2 + \kk + 2)
    -   2t_2f_A\kk
    +   2t_3(2rf'_A - 2f_A - \kk)                              \nn
  & - & 2t_6\kk(rf'_A - f_A - 1)
    -   2t_9\kk(rf'_A - f_A + 1)
    +   2p_1r^2H'_0                                            \nn
r^2t''_2 & + & 2rt'_2 - r^2\ddot{t}_2                          \nn
  & = & -\;2t_1f_A
    +   t_2(r^2H_0^2 + f_A^2 + \kk + 1)
    -   2t_3(rf'_A - f_A - 1)                                  \nn
  & + & 2t_6(rf'_A - f_A - \kk + 1)
    -   2t_9(rf'_A - f_A + 1)
    +   2p_2r^2H'_0                                            \nn
r^2t''_3 & + & 2rt'_3 - r^2\ddot{t}_3                          \nn
  & = & -\;2t_1
    -   2t_2(rf'_A - f_A + 1)
    +   t_3(r^2H_0^2 + 2f_A^2 + \kk)                           \nn
  & - & t_6[(3f_A - 1)(f_A - 1) + 2f_A\kk]
    +   t_9(3f_A - 1)(f_A - 1)                                 \nn
  & - & p_2rH_0(f_A - 1)                                       \nn
r^2t''_6 & + & 2rt'_6 - r^2\ddot{t}_6                          \nn
  & = & \{-\;2t_1(rf'_A - f_A + 1)
    -   2t_2(rf'_A - f_A + \kk + 1)                            \nn
  & - & t_3(f_A^2 + 2f_A\kk - 1)
    +   t_9[2\kk f_A(f_A - 1)
        + (3f_A - 1)(f_A - 1)]                                 \nn
  & + & t_6[\kk(r^2H_0^2 + f_A^2 + \kk - 1)
        - (3f_A - 1)(f_A - 1)]                                 \nn
  & - & p_1rH_0(f_A - 1)
    -   p_2rH_0(f_A - 1) \}/\kk                                \nn
r^2t''_9 & + & 2rt'_9 - r^2\ddot{t}_9                          \nn
  & = & \{-\;2t_1(rf'_A - f_A + 1)
    -   2t_2(rf'_A - f_A + 1)
    -   t_3(f_A^2 - 1)                                         \nn
  & + & t_6[2\kk f_A(f_A - 1)
        - (3f_A - 1)(f_A - 1)]                                 \nn
  & + & t_9[\kk(r^2H_0^2 + f_A^2 + \kk - 1)
        + (3f_A - 1)(f_A - 1)]                                 \nn
  & - & p_1rH_0(f_A - 1)
    -   p_2rH_0(f_A - 1) \}/\kk                                \nn
r^2p''_1 & + & 2rp'_1 - r^2\ddot{p}_1                          \nn
  & = & \{4t_1r^2H'_0
    +   4t_3rH_0(f_A - 1)
    -   2t_6rH_0\kk(f_A - 1)                                   \nn
  & - & 2t_9rH_0\kk(f_A - 1)
    -   2p_2\kk(f_A + 1)                                       \nn
  & + & p_1[r^2\xiq(H_0^2 - 1) + 2r^2H_0^2
        + (f_A + 1)^2 + 2\kk] \}/2                             \nn
r^2p''_2 & + & 2rp'_2 - r^2\ddot{p}_2                          \nn
  & = & \{4t_2r^2H'_0
    -   2t_3rH_0(f_A - 1)
    +   2t_6rH_0(f_A - 1)                                      \nn
  & - & 2t_9rH_0(f_A - 1)
    -   2p_1(f_A + 1)                                          \nn
  & + & p_2[r^2\xiq(H_0^2 - 1) + 2r^2H_0^2
        + f_A^2 + 2\kk - 1] \}/2
\end{eqnarray}
for the magnetic sector, by
\begin{eqnarray}
r^2t''_4 & + & 2rt'_4 - r^2\ddot{t}_4                          \nn
  & = & t_4(r^2H_0^2 + f_A^2 + \kk + 1)
    -   2t_5(rf'_A - f_A + 1)                                  \nn
  & + & 2t_7(rf'_A - f_A - \kk + 1)
    +   2t_8(rf'_A - f_A + 1)
    +   2p_3r^2H'_0                                            \nn
r^2t''_5 & + & 2rt'_5 - r^2\ddot{t}_5                          \nn
  & = & -\;2t_4(rf'_A - f_A + 1)
    +   t_5(r^2H_0^2 + 2f_A^2 + \kk)                           \nn
  & - & t_7(3f_A - 1)(f_A - 1)
    -   t_8[(3f_A - 1)(f_A - 1) + 2f_A\kk]                     \nn
  & - & p_3rH_0(f_A - 1)                                       \nn
r^2t''_7 & + & 2rt'_7 - r^2\ddot{t}_7                          \nn
  & = & \{2t_4(rf'_A - f_A - \kk + 1)
    +   t_5(f_A^2 - 1)                                         \nn
  & + & t_7[\kk(r^2H_0^2 + f_A^2 + \kk - 1)
        + (3f_A - 1)(f_A - 1)]                                 \nn
  & - & t_8[2\kk f_A(f_A - 1)
        - (3f_A - 1)(f_A - 1)]                                 \nn
  & + & p_3rH_0(f_A - 1)
    -   h_1rH_0(f_A - 1) \}/\kk                                \nn
r^2t''_8 & + & 2rt'_8 - r^2\ddot{t}_8                          \nn
  & = & \{-\;2t_4(rf'_A - f_A + 1)
    -   t_5(f_A^2 + 2f_A\kk - 1)                               \nn
  & - & t_7[2\kk f_A(f_A - 1)
        + (3f_A - 1)(f_A - 1)]                                 \nn
  & + & t_8[\kk(r^2H_0^2 + f_A^2 + \kk - 1)
        - (3f_A - 1)(f_A - 1)]                                 \nn
  & - & p_3rH_0(f_A - 1)
    +   h_1rH_0(f_A - 1) \}/\kk                                \nn
r^2p''_3 & + & 2rp'_3 - r^2\ddot{p}_3                          \nn
  & = & \{4t_4r^2H'_0
    -   2t_5rH_0(f_A - 1)
    +   2t_7rH_0(f_A - 1)
    +   2t_8rH_0(f_A - 1)                                      \nn
  & + & p_3[r^2\xiq(H_0^2 - 1) + 2r^2H_0^2
        + f_A^2 + 2\kk - 1]                                    \nn
  & + & 2h_1(f_A - 1) \}/2                                     \nn
r^2h''_1 & + & 2rh'_1 - r^2\ddot{h}_1                          \nn
  & = & \{-\;4t_5rH_0(f_A - 1)
    -   2t_7rH_0\kk(f_A - 1)
    +   2t_8rH_0\kk(f_A - 1)                                   \nn
  & + & 2p_3\kk(f_A - 1)
    +   h_1[r^2\xiq(3H_0^2 - 1)
        + (f_A - 1)^2 + 2\kk] \}/2
\end{eqnarray}
for the electric sector and by
\begin{eqnarray}
\label{dgls1}
r^2s''_1 & + & 2rs'_1 - r^2\ddot{s}_1                          \nn
  & = & s_1(r^2H_0^2 + 2f_A^2 + \kk)
    -   2s_2f_A\kk                                             \nn
\label{dgls2}
r^2s''_2 & + & 2rs'_2 - r^2\ddot{s}_2                          \nn
  & = & -\;2s_1f_A
    +   s_2(r^2H_0^2 + f_A^2 + \kk - 1)                        \nn
r^2s''_3 & + & 2rs'_3 - r^2\ddot{s}_3                          \nn
  & = & s_3(r^2H_0^2 + f_A^2 + \kk - 1)
\end{eqnarray}
for the Fadeev--Popov sector.

As one sees these equations do not have the form of Eq.
(\ref{flukdgl2}) with a symmetric potential $V_{ij}$. Such a form is
however needed on order to apply the Green function formalism in a
convenient way.

The transformations that bring the equations of motion
to a symmetric form have been found by an -- in fact very simple --
educated guessing. They are
\beq
 \begin{array}{rcl}
  t_1 & = & {\ds\frac{K(K+1)}{\sqrt{2K+1}}}
      \Big[{\ds\sqrt{\frac{(K+2)(K+1)}{2K+3}}}T_1
      - {\ds\frac{K}{\sqrt{2K-1}}} T_2                    \\[0.5em]
      &   & \hspace{5.5em} - {\ds\frac{K+1}{\sqrt{2K+3}}}T_3
      - {\ds\sqrt{\frac{K(K-1)}{2K-1}}} T_6\Big]                 \\
  t_2 & = & {\ds\frac{K}{\sqrt{(2K+1)(2K+3)}}}\Big[(K+1)T_3
        - \sqrt{(K+1)(K+2)}T_1\Big]                              \\
      & + & {\ds\frac{K+1}{\sqrt{(2K-1)(2K+1)}}}
      \Big[\sqrt{K(K-1)}T_6 - KT_2\Big]                          \\
  t_3 & = & {\ds\frac{K(K+1)}{\sqrt{(2K-1)(2K+1)}}}
            \Big[T_2+{\ds\sqrt{\frac{K}{K-1}}}T_6\Big]          \\
      & - & {\ds\frac{K(K+1)}{\sqrt{(2K+1)(2K+3)}}}
            \Big[T_3+{\ds\sqrt{\frac{K+1}{K+2}}}T_1\Big]         \\
  t_6 & = & {\ds\frac{K+1}{\sqrt{(2K+1)(2K-1)}}}
            \Big[T_2+{\ds\sqrt{\frac{K}{K-1}}}T_6\Big]          \\
      & + & {\ds\frac{K}{\sqrt{(2K+1)(2K+3)}}}
            \Big[T_3+{\ds\sqrt{\frac{K+1}{K+2}}}T_1\Big]   \\[0.5em]
  t_9 & = & T_9                                            \\[0.5em]
  p_1 & = & {\ds\frac{K(K+1)}{\sqrt{2K+1}}}
      \left[\sqrt{K}P_2-\sqrt{K+1}P_1\right]                     \\
  p_2 & = & {\ds\sqrt{\frac{K(K+1)}{2K+1}}}
            \left[\sqrt{K}P_1+\sqrt{K+1}P_2\right]
 \end{array}
\eeq
for the magnetic amplitudes and
\beq
 \begin{array}{rcl}
  t_4 & = & {\ds\frac{1}{\sqrt{2K+1}}}
  \Big[{\ds\frac{1}{\sqrt{K+1}}}T_5 -
  {\ds\frac{1}{\sqrt{K}}}T_4\Big]                                \\
  t_5 & = & {\ds\frac{1}{\sqrt{2K+1}}}\Big[{\ds\frac{1}
  {\sqrt{K-1}}}T_8 - {\ds\frac{1}{\sqrt{K+2}}}T_7\Big]           \\
  t_7 & = & {\ds\frac{1}{K(K+1)\sqrt{2K+1}}}
          \left[\sqrt{K}T_4+\sqrt{K+1}T_5\right]                 \\
  t_8 & = & {\ds\frac{1}{\sqrt{2K+1}}}
          \Big[{\ds\frac{1}{(K+1)\sqrt{K+2}}}T_7
        + {\ds\frac{1}{K\sqrt{K-1}}}T_8\Big]                     \\
  p_3 & = & {\ds \frac{1}{\sqrt{K(K+1)}}} P_3                    \\
  h_1 & = & H_1
 \end{array}
\eeq
for the electric amplitudes.
For the Fadeev--Popov amplitudes the transformations are
\beq
 \label{s1s2trafo}
 \begin{array}{rcl}
  s_1 & = & \sqrt{K}S_2 - \sqrt{K+1}S_1                          \\
  s_2 & = & {\ds\frac{1}{\sqrt{K+1}}}S_1
            -{\ds\frac{1}{\sqrt{K}}}S_2                          \\
  s_3 & = & S_3.                                                 \\
 \end{array}
\eeq

The n--tuples formed by the amplitudes of the various sectors
\begin{eqnarray}
  (\Psi_i^M) & = & (T_1,T_2,T_3,T_6,T_9,P_1,P_2)     \nn
  (\Psi_i^E) & = & (T_4,T_5,T_7,T_8,P_3,H_1)         \nn
  (\Psi_i^{FP}) & = & (S_1,S_2,S_3)
\end{eqnarray}
fulfill -- with the mass parameters $m_i$ and the angular momenta
$l_i$ from table \ref{koeffkg1} -- the required fluctuation
equations (\ref{flukdgl2}).

The symmetric potential of the fluctuation equations in the
form (\ref{flukdgl2}) has for the magnetic sector the
elements
\begin{eqnarray}
V^M_{11}
  & = & (H_0^2 - 1)
    -   \frac{4f'_A(K + 2)}{r(2K + 3)}                   \nn
  & + & \frac{f_A - 1}{r^2(2K + 3)}
        \left[4K^2 + 3(f_A + 5)K + 7f_A + 15\right]      \nn
V^M_{22}
  & = & (H_0^2 - 1)
    +   \frac{4f'_A(K - 1)}{r(2K - 1)}                   \nn
  & - & \frac{f_A - 1}{r^2K(4K^2 - 1}
        \left[8K^4 - 6(f_A - 1)K^3 + (5f_A - 7)K^2
        - 4K - 3f_A + 1\right]                           \nn
V^M_{33}
  & = & (H_0^2 - 1)
    +   \frac{4f'_A(K + 2)}{r(2K + 3)}
    +   \frac{f_A - 1}{r^2(2K + 3)(2K + 1)(K + 1)}       \nn
  &   & \times\left[8K^4 + 2(3f_A + 13)K^3 + 23(f_A + 1)K^2
        + 4(7f_A + 1)K + 8f_A\right]                     \nn
V^M_{44}
  & = & (H_0^2 - 1)
    -   \frac{4f'_A(K - 1)}{r(2K - 1)}                   \nn
  &   & - \frac{f_A - 1}{r^2(2K - 1)}
        \left[4K^2 - (3f_A + 7)K + 4(f_A + 1)\right]     \nn
V^M_{55}
  & = & (H_0^2 - 1)
    +   \frac{f_A - 1}{r^2K(K + 1)}
        \left[(f_A + 1)K(K + 1) + 3f_A - 1\right]        \nn
V^M_{66}
  & = & \frac{1}{2}(2 + \xiq)(H_0^2 - 1)
    +   \frac{f_A - 1}{2r^2}(2K + f_A + 3)               \nn
V^M_{77}
  & = & \frac{1}{2}(2 + \xiq)(H_0^2 - 1)
    -   \frac{f_A - 1}{2r^2}(2K - f_A - 1)               \nn
V^M_{12}
  & = & -\sqrt{\frac{(K + 1)(K + 2)(2K - 1)}{2K + 3}}
        \frac{(f_A - 1)^2}{r^2(2K + 1)}                  \nn
V^M_{13}
  & = & -\sqrt{\frac{K + 2}{K + 1}} \bigg\{
        \frac{2f'_A}{r(2K + 3)}                          \nn
  &   &  -\frac{f_A - 1}{r^2(2K + 3)(2K + 1)}
        \left[4(f_A + 1)K + f_A + 3\right]\bigg\}    \nn[0.5em]
V^M_{14} & = & 0                                         \nn
V^M_{15}
  & = & -\sqrt{\frac{K + 2}{(2K + 1)(2K + 3)(K + 1)}}    \nn
  &   &  \times\bigg\{\frac{2f'_AK}{r}
         - \frac{f_A - 1}{r^2}
         \left[2(f_A + 1)K - f_A +1\right] \bigg\}        \nn
V^M_{16}
  & = & - \sqrt{\frac{K + 2}{2K + 3}}\frac{1}{r}
        \left[2rH'_0 - (f_A - 1)H_0\right]                 \nn[0.5em]
V^M_{17} & = & 0                                           \nn
V^M_{23}
  & = & \frac{1}{\sqrt{(2K + 3)(2K - 1)}}
        \frac{2(f_A - 1)^2}{r^2(2K + 1)}                    \nn
V^M_{24}
  & = & \sqrt{\frac{K - 1}{K}} \left\{
        \frac{2f'_A}{r(2K - 1)}
    -   \frac{f_A - 1}{r^2(4K^2 - 1)}
        \left[4(f_A + 1)K + 3f_A + 1\right] \right\}         \nn
V^M_{25}
  & = & \frac{1}{\sqrt{4K^2 - 1}} \left\{
        \frac{2f'_A(K + 1)}{rK} \right.                  \nn
  &   &  + \left.\frac{f_A - 1}{r^2K} \left[2(f_A - 1)K^2
        + (f_A - 3)K - 3f_A + 1\right] \right\}          \nn[0.5em]
V^M_{26} & = & 0                                         \nn
V^M_{27}
  & = & -\frac{1}{\sqrt{K(2K - 1)}}\frac{1}{r}
        \left[2rH'_0K + (f_A - 1)H_0(K - 1)\right]       \nn
V^M_{34}
  & = & -\sqrt{\frac{K(K - 1)(2K + 3)}{2K - 1}}
        \frac{(f_A - 1)^2}{r^2(2K + 1)}                  \nn
V^M_{35}
  & = & \frac{1}{\sqrt{(2K + 1)(2K + 3)}}                \nn
  &   & \times\bigg\{\frac{2f'_AK}{r}
        + \frac{f_A - 1}{r^2(K + 1)}\left[2(f_A - 1)K^2
        + (3f_A - 1)K - 2f_A + 2\right]\bigg\}           \nn
V^M_{36}
  & = & \frac{1}{\sqrt{(K + 1)(2K + 3)}}
        \frac{1}{r} \left[2rH'_0(K + 1)
        + (f_A - 1)H_0(K + 2)\right]                     \nn[0.5em]
V^M_{37} & = & 0                                         \nn
V^M_{45}
  & = & -\sqrt{\frac{K - 1}{K(4K^2 - 1)}} \left\{
        \frac{2f'_A(K + 1)}{r} - \frac{f_A - 1}{r^2}
        \left[(f_A + 1)K + 3f_A + 1\right] \right\}      \nn[0.5em]
V^M_{46} & = & 0                                         \nn
V^M_{47}
  & = & \sqrt{\frac{K - 1}{2K - 1}}\frac{1}{r}
        \left[2rH'_0 - (f_A - 1)H_0\right]               \nn
V^M_{56}
  & = & \frac{K}{\sqrt{(K + 1)(2K + 1)}}
        \frac{1}{r} (f_A - 1)H_0                         \nn
V^M_{57}
  & = & -\frac{K + 1}{\sqrt{K(2K + 1)}} \frac{1}{r}
        (f_A - 1)H_0                                     \nn[0.5em]
V^M_{67} & = & 0.
\end{eqnarray}

For the electric sector we obtain
\begin{eqnarray}
V^E_{11}
  & = & (H_0^2 - 1)
    -   \frac{4f'_A}{r(2K + 1)}                          \nn
  &   &  + \frac{f_A - 1}{r^2(2K + 1)(K + 1)}
        \left[2(f_A + 1)K^2 + (3f_A + 7)K
        + 4f_A + 4\right]                                \nn
V^E_{22}
  & = & (H_0^2 - 1)
    +   \frac{4f'_A}{r(2K + 1)}                          \nn
  &   &  + \frac{f_A - 1}{r^2K(2K + 1)}
        \left[2(f_A + 1)K^2 + (f_A - 3)K
        + 3f_A - 1\right]                                \nn
V^E_{33}
  & = & (H_0^2 - 1)                                      \nn
  &   &  + \frac{f_A - 1}{r^2(2K + 1)(K + 1)}
        \left[4K^3 + (3f_A + 11)K^2
        + 9(f_A + 1)K + 3f_A + 3\right]                  \nn
V^E_{44}
  & = & (H_0^2 - 1)
    -   \frac{f_A - 1}{r^2K(2K + 1)}
        \left[4K^3 - (3f_A - 1)(K^2-K-1)\right]          \nn
V^E_{55}
  & = & \frac{1}{2}(\xiq + 2)(H_0^2 - 1)
    +   \frac{1}{2r^2}(f_A^2 - 1)                        \nn
V^E_{66}
  & = & \frac{3}{2}\xiq(H_0^2 - 1)
    +   \frac{1}{2r^2}(f_A - 1)^2                        \nn
V^E_{12}
  & = & - \frac{1}{\sqrt{K(K + 1)}}
        \frac{2rf'A - (f_A - 1)(3f_A + 1)}
        {r^2(2K + 1)}                                    \nn
V^E_{13}
 & = & - \frac{\sqrt{K(K + 2)}}{2K + 1}
        \left\{ \frac{2f'_A}{r}
        + \frac{f_A - 1}{r^2(K + 1)}
        \left[2(f_A - 1)K - f_A - 1\right] \right\}      \nn
V^E_{14}
  & = & \sqrt{\frac{K - 1}{K}} \frac{1}{r^2(2K + 1)}
        \{ 2rf'_A(K + 1)                                 \nn
  &   & - (f_A - 1)[2(f_A + 1)K + (3f_A + 1)]\}          \nn
V^E_{15}
  & = & - \frac{1}{\sqrt{(K + 1)(2K + 1)}}
        \frac{1}{r} \left[2rH'_0(K + 1) - (f_A - 1)H_0\right] \nn
V^E_{16}
  & = & - \sqrt{\frac{K}{2K + 1}}\frac{1}{r} (f_A - 1)H_0     \nn
V^E_{23}
  & = & \sqrt{\frac{K + 2}{K + 1}} \frac{1}{r^2(2K + 1)}
        \left\{2rf'_AK - (f_A - 1)
        \left[2(f_A + 1)K - f_A + 1\right]\right\}            \nn
V^E_{24}
  & = & - \frac{\sqrt{K^2 - 1}}{K(2K + 1)r^2}
        \left\{2rf'_AK + (f_A - 1)
        \left[2(f_A - 1)K + 3f_A - 1\right]\right\}           \nn
V^E_{25}
  & = & \frac{1}{\sqrt{K(2K + 1)}}\frac{1}{r}
        \left[2rH'_0K + (f_A - 1)H_0\right]                   \nn
V^E_{26}
  & = & - \sqrt{\frac{K + 1}{2K + 1}}\frac{1}{r} (f_A - 1)H_0 \nn
V^E_{34}
  & = & - \frac{\sqrt{(K - 1)(K + 2)}} {(2K + 1)r^2} (f_A - 1)^2 \nn
V^E_{35}
  & = & \sqrt{\frac{K(K + 2)}
        {(K + 1)(2K + 1)}}\frac{1}{r} (f_A - 1)H_0            \nn
V^E_{36}
  & = & \sqrt{\frac{K + 2}{2K + 1}} \frac{1}{r} (f_A - 1)H_0  \nn
V^E_{45}
  & = & - \sqrt{\frac{K^2 - 1}{K(2K + 1)}} \frac{1}{r}
        (f_A - 1)H_0                                          \nn
V^E_{46}
  & = & \sqrt{\frac{K - 1}{2K + 1}} \frac{1}{r} (f_A - 1)H_0  \nn
V^E_{56}
  & = & \frac{\sqrt{K(K + 1)}}{r^2} (f_A - 1)
\end{eqnarray}

and finally for the Fadeev--Popov sector
\begin{eqnarray}
\label{vfp11}
V^{FP}_{11}
  & = & (H_0^2 - 1) + \frac{f_A - 1}{(2K + 1)r^2}
        \left[4K^2 + (3f_A + 7)K + 2f_A + 2\right]            \nn
\label{vfp22}
V^{FP}_{22}
  & = & (H_0^2 - 1) - \frac{f_A - 1}{(2K + 1)r^2}
        \left[4K^2 - (3f_A - 1)K - f_A - 1\right]             \nn
V^{FP}_{33}
  & = & (H_0^2 - 1) + \frac{1}{r^2} (f_A^2 - 1)               \nn
V^{FP}_{12}
 & = & - \frac{\sqrt{K(K + 1)}} {(2K + 1)r^2} (f_A - 1)^2\nn[0.5em]
V^{FP}_{13} & = & 0                                      \nn[0.5em]
V^{FP}_{23} & = & 0.
\end{eqnarray}

We have again used REDUCE to obtain these potentials from the
equations of motion for the transformed amplitudes.

In the case $K=0$ the \opi-- and \opj--operators act on the constant
spherical harmonic $Y_{00}$. Therefore only the operators $\opi^1$,
$\opi^3$, $\opi^5$ and $\opj^1$ contribute to the fluctuation
Lagrangean. We can account for this by setting -- in the Lagrangean
-- the amplitudes $t_2$, $t_4$, $t_6$, $t_7$, $t_8$, $t_9$, $p_2$,
$p_3$, $s_2$ and $s_3$ equal to zero. For the remaining amplitudes
$t_1$, $t_3$, $t_5$, $p_1$, $h_1$ and  $s_1$ one obtains then the
Euler--Lagrange--equations
\begin{eqnarray}
r^2t''_1 & + & 2rt'_1 - r^2\ddot{t}_1                          \nn
  & = & t_1(r^2H_0^2 + 2f_A^2 + 2)
    +   4t_3(rf'_A - f_A)
    +   2p_1r^2H'_0                                            \nn
r^2t''_3 & + & 2rt'_3 - r^2\ddot{t}_3                          \nn
  & = & 2t_1(rf'_A - f_A)
    +   t_3(r^2H_0^2 + 3f_A^2 - 1)
    +   p_1rH_0(f_A - 1)                                       \nn
r^2t''_5 & + & 2rt'_5 - r^2\ddot{t}_5                          \nn
  & = & t_5(r^2H_0^2 + 3f_A^2 - 1)
    -   h_1rH_0(f_A - 1)                                       \nn
r^2p''_1 & + & 2rp'_1 -  r^2\ddot{p}_1                         \nn
  & = & \{4t_1r^2H'_0
    +   4t_3rH_0(f_A - 1)                                      \nn
  & + & p_1[r^2\xiq(H_0^2 - 1) + 2r^2H_0^2
        + (f_A + 1)^2]  \}/2                                   \nn
r^2h''_1 & + & 2rh'_1 - r^2\ddot{h}_1                          \nn
  & = & \{-4t_5rH_0(f_A - 1)
    +   h_1[r^2\xiq(3H_0^2 - 1) + (f_A - 1)^2] \}/2            \nn
r^2s''_1 & + & 2rs'_1 - r^2\ddot{s}_1                          \nn
  & = & s_1(r^2H_0^2 + 2f_A^2).
\end{eqnarray}

Changing the basis as
\beq
  \begin{array}{l@{\qquad}l@{\qquad}l}
    t_1 = \sqrt{2} T_3 - 2 T_1    &
    t_5 = \sqrt{2} T_5            &
    s_1 = S_1                     \\
    t_3 = \sqrt{2} T_3 + T_1      &
    h_1 = 2 H_1                   &\\
    p_1 = \sqrt{6} P_1            &&
  \end{array}
\eeq
and introducing the n-tuples
\beq
  (\eta_i^M) = (T_1,T_3,P_1), \quad
  (\eta_i^E) = (T_5,H_1) \quad\mbox{and}\quad
  \eta^{FP} = S_1
\eeq
for the fields the differential equations take the required form
of Eq. \ref{flukdgl2}. The masses $m_i$ and angular momenta $l_i$
are shown in table \ref{koeffk0}.

The potentials in the three sectors are
\begin{eqnarray}
V^M_{11} & = &
     (H_0^2 - 1) - \frac{1}{3r^2}
     \left[8rf'_A - (f_A - 1) (7f_A + 15)\right]               \nn
V^M_{22} & = &
     (H_0^2 - 1) + \frac{8}{3r^2}
     \left[rf'_A + (f_A - 1)f_A\right]                         \nn
V^M_{33} & = &
     \frac{1}{2} (\xiq + 2) (H_0^2 - 1)
   + \frac{1}{2r^2} (f_A - 1) (f_A + 3)                        \nn
V^M_{12} & = &
   - \frac{\sqrt{2}}{3r^2}
     \left[2rf'_A - (f_A - 1)(f_A + 3)\right]                  \nn
V^M_{13} & = &
     \sqrt{\frac{2}{3}} \frac{1}{r}
     \left[(f_A - 1)H_0 - 2rH'_0\right]                        \nn
V^M_{23} & = &
     \frac{2}{\sqrt{3}r} \left[(f_A - 1)H_0 + rH'_0\right]     \nn
V^E_{11} & = &
     (H_0^2 - 1) + \frac{3}{r^2} (f_A^2 - 1)                   \nn
V^E_{22} & = &
     \frac{3}{2}\xiq(H_0^2 - 1) + \frac{1}{2r^2} (f_A - 1)^2   \nn
V^E_{12} & = &
   - \frac{\sqrt{2}}{r}(f_A - 1)H_0                            \nn
V^{FP} & = &
     (H_0^2 - 1) + \frac{2}{r^2} (f_A^2 - 1).
\end{eqnarray}

Another exceptional case arises for  $K=1$. Here the \opj-- and
\opi--operators act on the spherical harmonics $Y_{1M} \propto
\hat{x}$. The action of $\opi^3$ becomes equal up to a sign to the
one of $\opi^6$ and similarly the one of $\opi^5$ to the one of
$\opi^8$ (see Eq. (\ref{dsfa})). Therefore a linear combination of
the amplitudes $t_3$ and $t_6$ and of $t_5$ and $t_8$ respectively
can be chosen to vanish. We have taken the choice
\beq
  t_6 = 0 \quad\mbox{and}\quad t_8=0.
\eeq
For the remaining 14 amplitudes we find
\begin{eqnarray}
r^2t''_1 & + & 2rt'_1 - r^2\ddot{t}_1                          \nn
   & = & t_1 (r^2 H_0^2 + 2 f_A^2 + 4)
     -   4 t_2 f_A
     +   4 t_3 (r f'_A - f_A - 1)                              \nn
   & - & 4 t_9 (r f'_A - f_A + 1)
     -   2 p_1 r^2 H'_0                                        \nn
r^2t''_2 & + & 2rt'_2 - r^2\ddot{t}_2                          \nn
   & = & - 2 t_1 f_A
     +   t_2 (r^2 H_0^2 + f_A^2 + 3)
     -   2 t_3 (r f'_A - f_A - 1)                              \nn
   & - & 2 t_9 (r f'_A - f_A + 1)
     +   2 p_2 r^2 H'_0                                         \nn
r^2t''_3 & + & 2rt'_3 - r^2\ddot{t}_3                          \nn
   & = & \{2 t_1 (r f'_A - f_A - 1)
     -   2 t_2 (r f'_A - f_A - 1)                              \nn
   & + & t_3 (2 r^2 H_0^2 + 5 f_A^2 + 4 f_A + 3)
     -   t_9 (f_A^2 - 1)
     +   p_1 r H_0 (f_A - 1)                                   \nn
   & - & p_2 r H_0 (f_A - 1) \}/2                              \nn
r^2t''_9 & + & 2rt'_9 - r^2\ddot{t}_9                          \nn
  & = &  \{- 2 t_1 (r f'_A - f_A + 1)
    -    2 t_2 (r f'_A - f_A + 1)
    -    t_3 (f_A^2 - 1)                                       \nn
  & + &  t_9 (2 r^2 H_0^2 + 5 f_A^2 - 4 f_A + 3)
    -    p_1 r H_0 (f_A - 1)                                   \nn
  & - &  p_2 r H_0 (f_A - 1) \} /2                             \nn
r^2p''_1 & + & 2rp'_1 - r^2\ddot{p}_1                          \nn
  & = &  \{4 t_1 r^2 H'_0
    +    4 t_3 r H_0 (f_A - 1)
    -    4 t_9 r H_0 (f_A - 1)                                 \nn
  & + &  p_1 [r^2 \xiq (H_0^2 - 1) + 2 r^2 H_0^2
         + f_A^2 + 2 f_A + 5]
    -    4 p_2 (f_A + 1) \} /2                                 \nn
r^2p''_2 & + & 2rp'_2 - r^2\ddot{p}_2                          \nn
  & = &  \{2 t_2 r^2 H'_0
    -    2 t_3 r H_0 (f_A - 1)
    -    2 t_9 r H_0 (f_A - 1)
    -    2 p_1 (f_A + 1)                                       \nn
  & + &  p_2 [r^2 \xiq (H_0^2 - 1) + 2 r^2 H_0^2 + f_A^2 + 3]
         \} /2
\end{eqnarray}
for the magnetic sector,
\begin{eqnarray}
r^2t''_4 & + & 2rt'_4 - r^2\ddot{t}_4                          \nn
  & = &  t_4 (r^2 H_0^2 + f_A^2 + 3)
    -    2 t_5 (r f'_A - f_A + 1)
    +    2 t_7 (r f'_A - f_A - 1)                              \nn
  & + &  2 p_3 r^2 H'_0                                        \nn
r^2t''_5 & + & 2rt'_5 - r^2\ddot{t}_5                          \nn
  & = &  \{- 2 t_4 (r f'_A - f_A + 1)
    +    t_5 (2 r^2 H_0^2 + 5 f_A^2 + 4 f_A + 3)               \nn
  & + &  t_7 (f_A^2 - 1)
    -    p_3 r H_0 (f_A - 1)
    -    h_1 r H_0 (f_A - 1) \} /2                             \nn
r^2t''_7 & + & 2rt'_7 - r^2\ddot{t}_7                          \nn
  & = &  \{ 2 t_4 (r f'_A - f_A - 1)
    +    t_5 (f_A^2 - 1)
    +    t_7 (2 r^2 H_0^2 + 5 f_A^2 - 4 f_A + 3)               \nn
  & + &  p_3 r H_0 (f_A - 1)
    -    h_1 r H_0 (f_A - 1) \} /2                             \nn
r^2p''_3 & + & 2rp'_3 - r^2\ddot{p}_3                          \nn
  & = &  \{ 4 t_4 r^2 H'_0
    -    2 t_5 r H_0 (f_A - 1)
    +    2 t_7 r H_0 (f_A - 1)                                 \nn
  & + &  p_3 [r^2 \xiq (H_0^2 - 1) + 2 r^2 H_0^2 + f_A^2 + 3]
    +    2 h_1 (f_A - 1) \} /2                                 \nn
r^2h''_1 & + & 2rh'_1 - r^2\ddot{h}_1                          \nn
  & = &  \{- 4 t_5 r H_0 (f_A - 1)
    -    4 t_7 r H_0 (f_A - 1)
    +    4 p_3 (f_A - 1)                                       \nn
  & + &  h_1 [r^2 \xiq (3 H_0^2 - 1) + f_A^2 - 2 f_A + 5] \} /2
\end{eqnarray}
for the electric sector and
\begin{eqnarray}
r^2s''_1 & + & 2rs'_1 - r^2\ddot{s}_1                          \nn
  & = &  s_1 (r^2 H_0^2 + 2 f_A^2 + 2)
    -    4 s_2 f_A                                             \nn
r^2s''_2 & + & 2rs'_2 - r^2\ddot{s}_2                          \nn
  & = &  - 2 s_1 f_A
    +    s_2 (r^2 H_0^2 + f_A^2 + 1)                           \nn
r^2s''_3 & + & 2rs'_3 - r^2\ddot{s}_3                          \nn
  & = &  s_3 (r^2 H_0^2 + f_A^2 + 1)
\end{eqnarray}
for the Fadeev--Popov sector. The amplitudes  are transformed as
\beq
 \begin{array}{l@{\qquad}l@{\quad}l}
  t_1={\ds\frac{2}{\sqrt{5}}}\left(2T_1+\sqrt{5}T_2+
  \sqrt{6}T_3\right)  &   t_4={\ds\frac{1}{\sqrt{3}}}
  \left(\sqrt{2}T_5-2T_4\right)              &
  s_1=S_2-\sqrt{2}S_1                                         \\
  t_2={\ds\frac{1}{\sqrt{5}}}\left(\sqrt{6}T_3+2\sqrt{5}T_2
  -2T_1\right) & t_5=T_7  &
  s_2=S_2+{\ds\frac{1}{\sqrt{2}}}S_1                            \\
  t_3={\ds\frac{1}{\sqrt{5}}}\left(\sqrt{6}T_3+3T_1\right)      &
  t_7={\ds\frac{1}{\sqrt{3}}}\left(T_4+\sqrt{2}T_5\right)       &
  s_3=S_3                                                       \\
  t_9=\sqrt{3}T_9                                               &
  p_3=\sqrt{2}P_3                                               &\\
  p_1=2\left(P_2-\sqrt{2}P_1\right)                             &
  h_1=2H_1                                                      &\\
  p_2=2P_2+\sqrt{2}P_1                                          &&
 \end{array}
\eeq
in order to get a symmetric potential and asymptotic decoupling.
With the n-tuples
\begin{eqnarray}
  (\eta_i^M) & = & (T_1,T_2,T_3,T_9,P_1,P_2)     \nn
  (\eta_i^E) & = & (T_4,T_5,T_7,P_3,H_1)         \nn
  (\eta_i^{FP}) & = & (S_1,S_2,S_3)
\end{eqnarray}
the fluctuation equations take the form of Eq. (\ref{flukdgl2}).
The masses $m_i$ and angular momenta $l_i$ are given in table
\ref{koeffk1}.

The elements of the symmetric potential are given by
\begin{eqnarray}
V^M_{11} & = &
   (H_0^2 - 1) + \frac{12 f'_A}{5r}
   + \frac{(f_A - 1)(65f_A + 61)}{30 r^2}                       \nn
V^M_{22} & = &
   (H_0^2 - 1) + \frac{4(f_A - 1)^2}{3r^2}                      \nn
V^M_{33} & = &
   (H_0^2 - 1) - \frac{12 f'_A}{5r}
   + \frac{(f_A - 1)(10 f_A + 34)}{5r^2}                        \nn
V^M_{44} & = &
   (H_0^2 - 1) + \frac{(f_A - 1)(5f_A + 1)}{2r^2}               \nn
V^M_{55} & = &
  \frac{1}{2}(2 + \xiq)(H_0^2 - 1)
  + \frac{(f_A - 1)(f_A + 5)}{2r^2}                             \nn
V^M_{66} & = &
  \frac{1}{2} (2 + \xiq)(H_0^2 - 1) + \frac{(f_A - 1)^2}{2r^2}  \nn
V^M_{12} & = &
  \frac{2(f_A - 1)^2}{3\sqrt{5}r^2}                             \nn
V^M_{13} & = &
   \frac{1}{5\sqrt{6}r^2}
   \left[(f_A - 1)(5f_A + 7) - 6rf'_A\right]                    \nn
V^M_{14} & = &
   - \frac{1}{2\sqrt{15}r^2}
   \left[(f_A - 1)(3f_A - 1) + 4rf'_A\right]                    \nn
V^M_{15} & = &
   - \frac{1}{\sqrt{10}r} \left[3(f_A - 1)H_0 + 4rH'_0\right]
   \nn[0.5em]
V^M_{16} & = & 0                                               \nn
V^M_{23} & = &
   - \sqrt{\frac{2}{15}} \frac{(f_A - 1)^2}{r^2}                \nn
V^M_{24} & = &
   \frac{4}{\sqrt{3}r^2} \left[(f_A - 1) - rf'_A\right]         \nn
V^M_{25} & = & 0                                                \nn
V^M_{26} & = &
   2 H'_0                                                       \nn
V^M_{34} & = &
   - \frac{1}{\sqrt{10}r^2}
   \left[(f_A - 1)(f_A + 3) - 2rf'_A\right]                     \nn
V^M_{35} & = &
   - \sqrt{\frac{3}{5}} \frac{1}{r}
   \left[(f_A - 1)H_0 - 2rH'_0\right]                    \nn[0.5em]
V^M_{36} & = & 0                                           \nn
V^M_{45} & = &
   \frac{1}{\sqrt{6}r} (f_A - 1)H_0                        \nn
V^M_{46} & = &
   - \frac{2}{\sqrt{3}r} (f_A - 1)H_0                    \nn[0.5em]
V^M_{56} & = & 0
\end{eqnarray}
for the magnetic sector, by
\begin{eqnarray}
V^E_{11} & = &
   (H_0^2 - 1) - \frac{4f'_A}{3r}
   + \frac{(f_A - 1)(9f_A + 13)}{6r^2}                     \nn
V^E_{22} & = &
   (H_0^2 - 1) + \frac{4f'_A}{3r}
   + \frac{(f_A - 1)(6f_A - 2)}{3r^2}                       \nn
V^E_{33} & = &
   (H_0^2 - 1) + \frac{(f_A - 1)(5f_A + 9)}{2r^2}           \nn
V^E_{44} & = &
   \frac{1}{2}(2 + \xiq)(H_0^2 - 1)
   + \frac{(f_A - 1)(f_A + 1)}{2r^2}                        \nn
V^E_{55} & = &
   \frac{3}{2}\xiq(H_0^2 - 1) + \frac{(f_A - 1)^2}{2r^2}    \nn
V^E_{12} & = &
   - \frac{1}{3\sqrt{2}r^2}
   \left[2rf'_A - (f_A - 1)(3f_A + 1)\right]                \nn
V^E_{13} & = &
   \frac{1}{2\sqrt{3}r^2}
   \left[4rf'_A + (f_A - 1)(f_A - 3)\right]                 \nn
V^E_{14} & = &
   -\frac{1}{\sqrt{6}r} \left[4rH'_0 - (f_A - 1)H_0\right]  \nn
V^E_{15} & = &
   - \frac{1}{\sqrt{3}r} (f_A - 1)H_0                       \nn
V^E_{23} & = &
   - \frac{1}{\sqrt{6}r^2}
   \left[2rf'_A - (f_A - 1)(f_A + 3)\right]                 \nn
V^E_{24} & = &
   \frac{1}{\sqrt{3}r} \left[2rH'_0 + (f_A - 1)H_0\right]   \nn
V^E_{25} & = &
   - \sqrt{\frac{2}{3}} \frac{1}{r} (f_A - 1) H_0           \nn
V^E_{34} & = &
   - \frac{1}{\sqrt{2}r} (f_A - 1) H_0                      \nn
V^E_{35} & = &
   - \frac{1}{r} (f_A - 1) H_0                              \nn
V^E_{45} & = &
   \frac{\sqrt{2}}{r^2} (f_A - 1)
\end{eqnarray}
for the electric sector and by
\begin{eqnarray}
V^{FP}_{11} & = &
   (H_0^2 - 1) + \frac{(f_A - 1)(5f_A + 13)}{3r^2}          \nn
V^{FP}_{22} & = &
  (H_0^2 - 1) + \frac{4(f_A - 1)^2}{3r^2}                   \nn
V^{FP}_{33} & = &
   (H_0^2 - 1) + \frac{(f_A - 1)(f_A + 1)}{r^2}             \nn
V^{FP}_{12} & = &
   -\frac{\sqrt{2}}{3r^2} (f_A - 1)^2                    \nn[0.5em]
V^{FP}_{13} & = & 0                                      \nn[0.5em]
V^{FP}_{23} & = & 0
\end{eqnarray}
for the Fadeev--Popov amplitudes.
\newpage
\setcounter{equation}{0}
\section{Zero Mode Prefactors}

The prefactors ${\cal N}_{tr}$ and ${\cal N}_{rot}$ are determined
\cite{La} by the normalization of the translation and rotation zero
modes. We have included  factors $1/2\pi$ which otherwise would
appear with the prefactor $T^{-3}$.
We have then (cf. \cite{ArMcL})
\begin{eqnarray}
{\cal N}_{tr}& =& N_{tr}^3                       \nn
{\cal N}_{rot} V_{rot} & = & 8 \pi^2 N_{rot}^3
\end{eqnarray}
where the normalization factors $N$ are given by
\beq
\frac{1}{2\pi} \int d^3x \psi_n^{tr,rot} \psi_n^{tr,rot}.
\eeq
The $\psi_n$ are the zero mode wave functions, the different
field components are assumed to have canonical normalization
(i.e. appearing as $\frac{1}{2}\partial_{\mu}\psi^\dagger_n
\partial^{\mu}\psi_n$
for each real component in the Lagrangean density).
The rotation and translation modes have been determined explicitly
in \cite{BaLa} for the sphaleron solution in the form (\ref{clans}).
The zero mode amplitudes were found to be
\bea
t_4 &=& -\frac{f_A-1}{r^2}                   \nn
t_5 &=& \frac{f_A+1}{r^2} -\frac{f'_A}{2r}   \nn
t_7 &=& \frac{-f'_A}{2r}                     \nn
h_1 &=&  H'_0
\eea
for the translation mode and
\beq
t_9 = \frac{f_A-1}{r}
\eeq
for the rotation mode.

For proper normalization (which was irrelevant in \cite{BaLa}) all
these contributions have to be multiplied by $\sqrt{4\pi/3}$ which
comes from the different normalization of the $Y_{1M}$ and the
$\hat{x}_M$ used in that calculation. Also the translation mode
amplitudes have to be multiplied by $M_W$ if they are generated
by the ordinary gradient, i.e. the derivatives w. r. t.
$\bar{x}_\mu$. We will include these additional factors at the end
(see Eqs. (\ref{normint}) and (\ref{zeroint})).

In this form the modes are not normalizable and do not satisfy the
background gauge condition (\ref{zwang}). The general form of the
infinitesimal gauge transformations (which also applies for
finite ones since we have expanded the fields linearly around the
classical solution) has also been given in \cite{BaLa}. For the
$K = 1$ channel it reads:
\bea
   \delta t_1 &=& g'_1                                           \nn
   \delta t_2 &=& g'_2                                           \nn
   \delta t_3 &=& \frac{g_1+g_2}{r}-\frac{(f_A-1)}{2r}(g_1+3g_2) \nn
   \delta t_4 &=& g'_3                                           \nn
   \delta t_5 &=& -\frac{3}{2}\,\frac{f_A-1}{r}g_3               \nn
   \delta t_7 &=& \frac{g_3}{r}+\frac{f_A-1}{2r}g_3              \nn
   \delta t_9 &=& -\frac{f_A-1}{2r}(g_1+g_2)                     \nn
   \delta p_1 &=& H_0 g_1/2                                      \nn
   \delta p_2 &=& H_0 g_2/2                                      \nn
   \delta p_3 &=& H_0 g_3/2                                      \nn
   \delta h_1 &=& H'_0.
\eea
The background gauge conditions read in terms of the partial
wave amplitudes
\bea
r t'_1 + 2t_1 &=& 2 (f_A+1) t_3 -2  (f_A-1) t_9 + 2  r H_0 p_1   \nn
r t'_2 + 2t_2 &=& - (f_A+1) t_3 -  (f_A-1) t_9 + 2   r H_0 p_2   \nn
r t'_4 + 2t_4 &=& - (f_A-1) t_5 + (f_A+1) t_7 +  2  r H_0 p_3.
\eea

Inserting the amplitudes and gauge functions leads to three
differential equations for the gauge functions
\bea
  g_1'' + \frac{2}{r}g_1' & = & \left[ H_0^2 + \frac{2(f_A^2+1)}
  {r^2} \right]g_1-\frac{4f_A}{r^2}g_2-\frac{2(f_A-1)^2}{r^2}    \nn
  g_2'' + \frac{2}{r}g_2' & = & \left[ H_0^2 + \frac{f_A^2+1}{r^2}
   \right]g_2-\frac{2f_A}{r^2}g_1-\frac{(f_A-1)^2}{r^2}        \nn
  g_3'' + \frac{2}{r}g_3' & = & \left[ H_0^2 + \frac{f_A^2+1}{r^2}
            \right]g_3-\frac{(f_A-1)^2}{r^3}
\eea
which have to be solved with boundary conditions that
ensure the normalizability of the zero modes.
The solution for $g_3$ can be found explicitly:
\beq
  g_3 = \frac{1-f_A}{r}
\eeq
so that the translation mode becomes
\bea
  t_4 & = & -\frac{f'_A}{r}                        \nn
  t_5 & = & \frac{f_A^2-1}{2r^2}-\frac{f'_A}{2r}   \nn
  t_7 & = & -\frac{f_A^2-1}{2r^2}-\frac{f'_A}{2r}  \nn
  p_3 & = & -\frac{f_A-1)H_0}{2r}                  \nn
  h_1 & = & H'_0.
\eea
The gauge transformed rotation mode becomes
\bea
  t_1 & = & g'_1                                         \nn
  t_2 & = & g'_2                                         \nn
  t_3 & = & \frac{g_1-g_2}{r}+\frac{f_A-1}{2r}(g_1-g_2)  \nn
  t_9 & = & -\frac{f_A-1}{2r}(g_1+g_2)+\frac{f_A-1}{r}   \nn
  p_1 & = & \frac{1}{2}H_0 g_1                           \nn
  p_2 & = & \frac{1}{2}H_0 g_2.
\eea
We have not been able, however, to solve the equations for the
gauge functions $g_1$ and $g_2$ analytically.
The normalization integrals $N^{tr,rot}$ are obtained by inserting
these amplitudes into the general expression
amplitudes as
\beq
\label{normint}
N^2 = \frac{1}{2\pi} \frac{4\pi}{3} \frac{1}{M_W g^2}
\int_0^\infty dr\,r^2 [t_1^2+2(t_2^2+t_4^2)+
4(t_3^2+t_5^2+t_7^2+t_9^2)+4 (h_1^2+p_1^2+2(p_2^2+p_3^2))].
\eeq
Here the first factor in front of the integral is a factor
``borrowed'' from the factors $\sqrt{2\pi T}$ which arise
for each extracted zero mode and which we have included
here as it was done in \cite{ArMcL}. The factor $4\pi/3$
has been explained above.
Finally the factors in front
of the different amplitudes come additionally from the normalization
of the tensors used in the expansion. Explicitly we obtain

\bea
\label{zeroint}
  N_{tr}^2 & = & \frac{8 M_W}{3g^2} \int_0^\infty dr
      \left[{f'_A}^2+\frac{(f_A^2-1)^2}{2r^2}+r^2{H'_0}^2
      +\frac{1}{2}H_0^2(f_A-1)^2 \right]                   \nn
  N_{rot}^2 & = & \frac{4}{3 M_W g^2} \int_0^\infty dr\,r^2 \left\{
       \frac{1}{2}{g'_1}^2+{g'_2}^2+2\left[\frac{g_1-g_2}{r}
       +\frac{f_A-1}{2r}(g_1-g_2)\right]^2 \right.               \nn
 & &   \left.+2\left[\frac{f_A-1}{r}
       -\frac{f_A-1}{2r}(g_1+g_2)\right]^2+\frac{1}{2}H_0^2g_1^2
       +H_0^2g_2^2\right\}.
\eea

The expression obtained for the translation mode can be
shown to be
\beq
N_{tr}^2= E_{cl}/2\pi
\eeq
as expected from a general virial theorem.
The rotation mode normalization should be related,
by a similar virial theorem, to the moment
of inertia of the sphaleron. We have evaluated the
normalization integrals numerically; we find
after taking into account factors $2^{\pm 3/2}$ due to
the different scales $M_W$ and $gv$ used
in the two publications -- the same results as \cite{cmcl1}
within the numerical accuracy, though
we have used a different gauge for the classical solution. Note that
the scale factors cancel in the product ${\cal N}_{rot}{\cal N}_
{trans}$.
\newpage
\setcounter{equation}{0}
\section{Determination of the Amplitudes ${h_n^\alpha}^\pm$}
\label{nystanh}
In this Appendix we will discuss briefly the
numerical evaluation of the functions ${h_n^\alpha}^\pm$,
i.e. the solutions of the differential equations (\ref{hdgl}).

The numerical integration of the differential equations (\ref{hdgl})
is started, for ${h_n^\alpha}^+$, at  some sufficiently high
$r=r_{max}$ with the initial condition $h_n^{\alpha +}(r_{max})=0$.

Starting the functions ${h_n^\alpha}^-$ at $r =0$ is by far more
difficult: the behaviour of these functions as $r  \to 0$
has to be determined analytically. This means that all the functions
that enter the differential equations
(\ref{hdgl}), i.e. the Bessel functions and their derivatives,
the classical profiles $f_A$ and $H_0$ and the
solutions ${h_n^\alpha}^-$  have to be expanded, for $r\to 0$
into Taylor series. As a first step which proves to be nontrivial
one has to find the leading behaviour of the solutions, since the
centrifugal barrier at $r=0$  differs from the
vacuum sector one.
If we write the leading behaviour as $r^\Delta$ we find
\beq
  \Delta = \left\{
    \begin{array}{cl}
       2 & \mbox{for the amplitude}\;S_3            \\
       1 & \mbox{for}\;P_2                          \\
       0 & \mbox{for}\;T_5,T_6,T_8,S_2              \\
      -1 & \mbox{for}\;P_1,P_3,H_1                  \\
      -2 & \mbox{for}\;T_1,T_2,T_3,T_4,T_7,T_9,S_1.
    \end{array}
  \right.
\eeq
With these parameters one obtains in each  $n \times n$ sector a
set of recursion relations for the next--to--leading
coefficients and a set of
starting conditions for $n$ independent solutions
labelled by $\alpha  =1,n$.

This expansion which has been determined up to the
second nonleading order in $r^2$ is used up to some
suitable $r =r_{min}$ at which the Nystr\"om integration is
then started.
The solutions found in this way do not yet satisfy the
boundary condition $h_n^{\alpha -} \to 0$ as $r \to \infty$.
However a set of such solutions can now be found
by a simple linear transformation.

A good check on the accuracy of the numerical
integration consists in checking the constancy of
the product $r^2 W^{\alpha\beta}(r)$
related to the Wronskian (see Eq. (\ref{wronf})).
For
\beq
  \label{rregion}
  r>\left\{
    \begin{array}{cl}
      5\cdot 10^{-3} & \mbox{for the amplitude}\;S_3  \\
      0.1            & \mbox{for all other sectors}
    \end{array}
 \right.
\eeq
this expression was found to be constant to 5 significant digits.
For smaller $r$ the numerical integration becomes delicate for all
sectors except the amplitude $S_3$, since some of the amplitudes
become singular as $r \to 0$. In this region we used the known
leading behaviour of the exact Green function
\beq
  r^2 G_{ii}(r,r,\nu) \propto \left\{
    \begin{array}{cl}
      r^3 & \mbox{for the amplitude}\;S_3  \\
      r   & \mbox{for all other sectors}
    \end{array}
  \right.
\eeq
with coefficients determined from the numerical results in the
reliable region (\ref{rregion}).
\end{appendix}

\section*{Figure Captions}
\parindent0pt

{\bf Fig. 1} The classical sphaleron energy $E_{cl}$:   \\
The figure shows the classical sphaleron energy $E_{cl}$ as
a function of $\xi=M_H/M_W$ in units of $M_W(T)/\alpha_w$.
\\
\\
{\bf Fig. 2} The zero--mode normalization factors:  \\
The solid line shows the normalization factor ${\cal N}_{tr}$ of the
translation mode and the dashed line the normalization factor
${\cal N}_{rot}$ of the rotation mode as a function of
$\xi=M_H/M_W$. The units are $(M_W/g^2)^{3/2}$ and
$(M_W g^2)^{-3/2 }$ respectively (see Eqs.(\ref{zeroint})).
\\
\\
{\bf Fig. 3}  The convergence of the K summation: \\
We show the partial sums $F(K_{max},\nu)$ as defined in
the text a function of $K_{max}$ for different values of $\nu$.
The dashed lines are the values obtained by including the sum
from $(K_{max}+1)$ to $\infty$ using the fit of Eq. (\ref{asfit}).
These values are seen to become independent of $K_{max}$ already
around $K_{max} \approx 10$.
\\
\\
{\bf Fig. 4} The function $F(\nu)$ for $\xi = M_H/M_W = 1$:  \\
The solid line shows $\nu F_{ren}(\nu)$, the dashed line the
unrenormalized $\nu F(\nu)$. The pole contribution of the unstable
mode has been removed (see Eq. (\ref{instmod})). The dotted lines
show the expected power behaviours at small and large $\nu$ and
the dash--dotted line the analytically known
tadpole contribution to $\nu F(\nu)$.
\\
\\
{\bf Fig. 5} The fluctuation determinant:   \\
The circles are our results, the crosses those of
CLMW. The full line is the estimate of Carson and
McLerran based on the DPY approximation and the
dashed line a perturbative estimate.
\newpage
\section*{Tables}

\begin{table}[h]
 \[
  \begin{array}[t]{|c||c|c|c|c|c|c|c|} \hline
    \xi & 0.4 & 0.5 & 0.6 & 0.8 & 1.0 & 1.5 & 2.0 \\  \hline
\omega_- & 1.32 & 1.36 & 1.39 & 1.45 & 1.51 & 1.62 & 1.71 \\ \hline
\ln\kappa & 6.18 & 5.89 & 5.68 & 5.50 & 5.48 & 5.62 & 5.71 \\ \hline
  \end{array}  \]
\caption{\label{restab} The results for $\ln \kappa$ for various
values  of $\xi=M_H/M_W$ together
   with  the frequencies  $\omega_{-}$ of the unstable mode}
\end{table}

\begin{table}[h]
\[
  \begin{array}{*{3}{c@{\qquad}}}
    {\begin{array}[t]{|*{3}{c|}}  \hline
      M & m_i & l_i               \\ \hline\hline
      1 & 1 & K + 2               \\ \hline
      2 & 1 &   K                 \\ \hline
      3 & 1 &   K                 \\ \hline
      4 & 1 & K - 2               \\ \hline
      5 & 1 &   K                 \\ \hline
      6 & 1 & K + 1               \\ \hline
      7 & 1 & K - 1               \\ \hline
    \end{array}} &
    {\begin{array}[t]{|*{3}{c|}}  \hline
      E & m_i & l_i               \\ \hline\hline
      1 &  1  & K + 1             \\ \hline
      2 &  1  & K - 1             \\ \hline
      3 &  1  & K + 1             \\ \hline
      4 &  1  & K - 1             \\ \hline
      5 &  1  &   K               \\ \hline
      6 & \xi &   K               \\ \hline
    \end{array}} &
    {\begin{array}[t]{|*{3}{c|}}  \hline
      FP & m_i & l_i              \\ \hline\hline
      1 & 1 & K + 1               \\ \hline
      2 & 1 & K - 1               \\ \hline
      3 & 1 &   K                 \\ \hline
    \end{array}}
  \end{array} \]
  \caption{\label{koeffkg1} Masses and Angular Momenta of the
  Amplitudes for $K>1$}
\end{table}

\begin{table}[h] \[
  \begin{array}{*{3}{c@{\qquad}}}
    {\begin{array}[t]{|*{3}{c|}} \hline
      M & m_i & l_i              \\ \hline\hline
      1 & 1 & 2                  \\ \hline
      2 & 1 & 0                  \\ \hline
      3 & 1 & 1                  \\ \hline
    \end{array}} &
    {\begin{array}[t]{|*{3}{c|}} \hline
      E & m_i & l_i              \\ \hline\hline
      1 &  1  & 1                \\ \hline
      2 & \xi & 0                \\ \hline
    \end{array}} &
    {\begin{array}[t]{|*{3}{c|}} \hline
      FP & m_i & l_i             \\ \hline\hline
      & 1 & 1                    \\ \hline
    \end{array}}
  \end{array} \]
  \caption{\label{koeffk0} Masses and Angular Momenta of the
  Amplitudes   for $K=0$}
\end{table}

\begin{table}[h] \[
  \begin{array}{*{3}{c@{\qquad}}}
    {\begin{array}[t]{|*{3}{c|}} \hline
      M & m_i & l_i              \\ \hline\hline
      1 & 1 & 1                  \\ \hline
      2 & 1 & 1                  \\ \hline
      3 & 1 & 3                  \\ \hline
      4 & 1 & 1                  \\ \hline
      5 & 1 & 2                  \\ \hline
      6 & 1 & 0                  \\ \hline
    \end{array}} &
    {\begin{array}[t]{|*{3}{c|}} \hline
      E & m_i & l_i              \\ \hline\hline
      1 &  1  & 2                \\ \hline
      2 &  1  & 0                \\ \hline
      3 &  1  & 2                \\ \hline
      4 &  1  & 1                \\ \hline
      5 & \xi & 1                \\ \hline
    \end{array}} &
    {\begin{array}[t]{|*{3}{c|}} \hline
      FP & m_i & l_i             \\ \hline\hline
      1 & 1 & 2                  \\ \hline
      2 & 1 & 0                  \\ \hline
      3 & 1 & 1                  \\ \hline
    \end{array}}
  \end{array} \]
  \caption{\label{koeffk1} Masses und Angular Momenta of the
  Amplitudes for $K=1$}
\end{table}

\end{document}